\documentclass[12pt]{article}
\usepackage[utf8]{inputenc}
\usepackage[T1]{fontenc}
\usepackage{mathtools}
\usepackage{ragged2e}
\usepackage{tabularx}
\usepackage{eurosym}
\usepackage[english]{babel}
\usepackage[margin=1in]{geometry}
\usepackage{amsmath,amsthm,amssymb}
\usepackage{indentfirst}
\usepackage{hyperref}
\usepackage{ulem}
\usepackage{setspace}
\usepackage{adjustbox}
\usepackage{rotating}       % for sidewaystable
\usepackage{threeparttable} % for nice table notes
\usepackage{subcaption}         % for subfloat
\usepackage{booktabs}       % for improved rules
\usepackage{array}
\usepackage{float} 
\usepackage[]{caption}
\usepackage[justification=justified,singlelinecheck=false, font={small},labelfont={bf}]{caption}
\usepackage{booktabs}    % For professional-looking rules
\usepackage{siunitx}     % For numeric alignment
\usepackage{subcaption}
\newcommand{\setfoot}[2]{%
    \footnote{#2}%
    \newcounter{#1}%
    \setcounter{#1}{\value{footnote}}%
}

\newcommand{\qedwhite}{\hfill \ensuremath{\Box}}

\newcommand\numeq[1]%
  {\stackrel{\scriptscriptstyle(\mkern-1.5mu#1\mkern-1.5mu)}{=}}
  
\title{Endogenous Product Design: A Linear Demand Approach}
\author{Afonso Rodrigues\setfoot{unia}{Department of Economics, University of Oxford (afonso.rodrigues@economics.ox.ac.uk). I have no relevant or material financial interests that relate to the research described in this paper. I gratefully acknowledge financial support by the Department of Economics at the University of Oxford and the Fundação para a Ciência e Tecnologia of the Ministry of Education, Science and Innovation (Portugal). I am indebted to my dissertation advisor Alexei Parakhonyak, as well as Bruno Pellegrino and Daniel Voelkening for their helpful insights.}}

\date{\today}

\begin{document}
\maketitle
\small

\begin{center}
    \textbf{Abstract}\\
\end{center}

This paper introduces a novel characteristics-based specification for linear demand to investigate endogenous product design. Characteristics are allowed to affect both consumers' product valuations and to what extent these compete. I demonstrate how such a specification can help linear demand deliver fresh insights in how firms may optimally design existing goods, as well as predict demand for new products. The framework is novel in its broad applicability to settings with any finite number of goods, firms, and product characteristics, with both vertical and horizontal differentiation across different market structures, and under asymmetry.\\

\textbf{JEL}: D21, D85, L20.\\

\textbf{Keywords}: linear demand, horizontal competition, vertical competition, product design, network formation, games on networks, monopoly, oligopoly, product differentiation.\\

\normalsize
%\doublespacing
\newpage
\section{Introduction}

Consider the following setup. A firm wishes to maximise profits by strategically setting the price(s) and characteristics of its good(s) within a product category. Consumers derive utility from said good(s) based on those characteristics; the product design choices of the firm's competitors also shape demand elasticities and the intensity of price competition. We want to know how said firm may optimise their good(s)' characteristics and what product differentiation and market structure are likely to arise as a result. How can we determine this?\\

An adequate modelling of endogenous product design is of broad interest. In industry, firms may benefit from a systematic understanding of design for strategic decision-making. Policy-makers and regulator regularly consider and apply minimum standard requirements, e.g. right-to-repair laws and the the USB-C standard in consumer electronics, or minimum free allowances for cabin bags in flights. A recent resurgence in interest in these types of requirements (for an European perspective, see Zeitlin and Rangoni, 2025) has further elevated the need for adequate analytical methods to consider their implications for firm behaviour.\\

In this paper, I formalise a theory of endogenous product design through a \textit{characteristics-based specification for linear demand}. The proposed approach is applied to analyse firm behaviour in differentiated products markets with a finite number of goods, firms, and product characteristics, in a representative consumer specification. To adjust their product(s)' horizontal and vertical differentiation relative to others, firms may design their goods both on the extensive side (i.e. which characteristics to include in their good, the \textit{orientation}), and the intensive side (i.e. how much of each to include, the \textit{intensity}). Characteristics differ in their utility but also in their \textit{salience} to product market competition, and the two need not be related: the color on a phone may be low utility but high salience; data privacy may be high utility but low salience. Characteristics may also impact substitution/complementarity between goods through both negative and positive interactions with each other: e.g. comfort and horsepower in cars on the one hand, sweetness and saltiness on popcorn on the other. The characteristics firms optimise over need not be independent of each other.\\

At the heart of the proposed characteristics-based linear demand approach sits an orthonormalised representation of product characteristics, which I rename as \textit{attributes}. Attributes provide the necessary tractability for a general model of product design. The key technical step is then a general approximation of the Hessian matrix of the representative consumer's utility function. In linear demand, the Hessian matrix, which reflects diminishing marginal utility due to satiation, is proportional to the price effects. I propose an approximation assuming \textit{only} that said matrix is at least partially determined by observed product characteristics. The approximation keeps exactly the structure that is identified by these, without imposing extra structure on what we do not observe.\\

I consider two market structures: multi-product monopoly and single-product oligopoly. I analyse each under different extents of control over their own product characteristics, as well as both where firm- and product-level characteristic costs do and do not differ, introducing asymmetries. Characteristics costs are used to reflect how increasing the intensity of a given characteristic often takes engineering effort, better inputs, or foregone room for other features; each increment gets harder as the easy improvements are exhausted first. \\

The model yields fresh insights on firm behaviour, remaining tractable even in these very general conditions. Under monopoly pricing, I show the model has a unique equilibrium strategy with no incentive for horizontal differentiation in observed characteristics. The monopolist prefers attribute alignment across all its goods. Cannibalisation alone is not a sufficiently strong incentive to diversify the product range; instead, the monopolist avoids spreading product quality investments across different directions and instead pushes the whole product line towards the direction yielding the greatest marginal utility per unit of cost to the average consumer. The monopolist is otherwise indifferent over the extent of vertical differentiation across (however many) goods, once aggregate attribute intensity is fixed. Under oligopoly pricing, I prove the existence of a unique symmetric equilibrium in product characteristics, without either horizontal or vertical differentiation in observed attributes. Competing firms do however distribute their product design across attributes more evenly. Competition makes firms more balanced in the attributes they load to minimise business stealing by other market participants. They increasingly emphasize the attributes with lesser competitive salience as the number of competitors increases.\\

Product-level asymmetry in the marginal cost of characteristics in a monopoly can result from differences in e.g. manufacturing technology. Under monopoly pricing, asymmetry breaks the indifference in how attributes are distributed across goods. The monopoly instead loads all attributes onto a single good: the good with the lowest costs per marginal utility gained across all attributes. A single good is shown to be sufficient to extract surplus from the consumer, despite complementarity between goods not being ruled out ex-ante. Under oligopoly pricing, firm-level cost asymmetry sustains horizontal differentiation. While a closed-form solution is not achieved, I show the model remains sufficiently nimble to converge numerically. \\

I extend the analysis to address demand prediction for new goods. I find that greater similarity between incumbents and entrants lowers both their demand and profits and discourages entry. If entry takes place, it shifts incumbent design in the exact opposite direction from the entrant’s. This suggests a potential strength in niche designs as an early strategy for new entrants, with a greater focus on a broader audience over time.\\

The remainder of the paper is organised as follows. Section 2 reviews the relevant literature. Section 3 sets out the theoretical framework and develops the characteristics-based representation of linear demand used throughout the paper. Section 4 studies optimal product design under monopoly pricing, and Section 5 under single-product oligopoly pricing. Section 6 extends the analysis to firm- and product-level heterogeneity in attribute costs. Section 7 examines product entry and its implications for market outcomes and post-entry design incentives. Section 8 presents a numerical example to illustrate the model’s mechanisms. Section 9 concludes.\\

\section{Literature Review}

This paper relates to growing interest in endogenous product design modelling.\\

Lancaster (1966) pioneered characteristics models in Economics, whereby a product is defined as a vector whose dimensions reflect different abstract characteristics, and consumers have preferences defined over said characteristics, purchasing products given their respective characteristic "basket". The earliest research on endogenous product design derives directly from that seminal contribution. Leland (1977) models product quality as a scalar design variable that changes the bundle of characteristics embodied in each unit of the good. The firm’s profit-maximising quality choice works through how a change in  alters the product’s characteristics and hence the price consumers pay. Drèze and Hagen (1978) models each product’s quality as a vector of objectively measurable characteristics per unit of output, and compares the profit-maximising product design to the Pareto-optimal. The two papers make generalisable statements regarding the monetary valuation of quality changes and implications for market efficiency, respectively.\\

Johnson and Myatt (2006) is a more recent major contribution to the literature, allowing further insights at a cost of greater model specification. That study analyses how changes in the distribution of consumer preferences, which can be triggered by the firm(s), affect product demand and profitability. Johnson and Myatt (2006) argues that firms find the most success by pursuing either a mass-market strategy, with low dispersion or a niche strategy, with high dispersion. Bar-Isaac, Caruana, and Cunat (2012) extend this analysis to a full competitive environment with a consumer search model, while Bar-Isaac, Caruana, and Cunat (2023) allows for marginal cost asymmetries.\footnote{Their definition of a consumer's valuation for a product matches surprisingly well with my approach. Linear demand is commonly derived from some version of a quasi-linear quadratic utility function (for more on this, see Amir, Erickson, and Jin, 2017; Choné and Linnemer, 2020). Quasi-linear quadratic utility is often interpreted from a mean-variance perspective, with the Hessian matrix defined as the variance-covariance matrix of the demand for goods (a common interpretation in Finance, see e.g. Campbell and Viceira, 2002, Ch. 2; Koijen and Yogo, 2019). In Johnson and Myatt (2006), the variance-covariance matrix of consumer preferences is also a function of weights placed on different product characteristics.} However, many of the results in the so-called \textit{targetted} design literature rely on an assumption that firms control not just their product characteristics, but the full dispersion of preferences. This dispersion is a single dimension, which drives all results regardless of the exact characteristics mix; different designs are treated as equivalent if they induce the same valuation dispersion.\\

Linear demand as a continuous demand system allows horizontal differentiation in a manner not captured solely by a product's appeal to a niche versus mass audience: goods can differentiate without losing broad interest. The \textit{generalised hedonic-linear demand} model of Pellegrino (2025), while not developed originally for the purpose, offers a promising way to leverage linear demand to study characteristics choice. Pellegrino (2025) studies market power by embedding Cournot oligopolistic competition in a general-equilibrium model calibrated to firm-level data for the universe of US public firms. Crucially, firms compete in a network of product-market rivalries defined by the \textit{cosine similarity} of their product characteristics bundle. The strength of the approach is in defining a specification for linear demand sufficiently tractable to enable multi-objective optimisation in characteristics. This is first noted in Voelkening (2026), which builds on Pellegrino (2025) for that purpose. The author studies welfare implications under a symmetric Cournot duopoly, where firms choose product characteristics prior to output. More recently, Miyashita (2026) extends this idea further, to multi-product monopoly and general oligopoly settings, with simultaneous characteristics and output choice. \\

The present paper contributes to this most recent spurt of the literature, while departing from it in several meaningful ways. I analyse Bertrand price competition à la Ushchev and Zenou (2018) and focus on firm behaviour under a broader set of scenarios, including asymmetries and firm entry. Similar to Voelkening (2026), I model a two-stage game, with product design followed by price setting. My model includes latent product characteristics and demand shocks; Miyashita (2026) also includes product-specific characteristics, but as public information prior to product design. Most significantly, I propose an alternative to cosine similarity for the purpose of endogenous product design. As a measure of competitive pressure, cosine similarity is insufficiently general: it lacks a micro-foundation; ascribes equal weights to all characteristics regardless of their salience to consumers; only accounts for observed characteristics; and does not address the overlapping information characteristics may contain for defining a given product. In practice, cosine similarity also excludes competition through vertical differentiation, by unit-normalising the bundles of characteristics of each good.\\

More broadly, the present paper makes a contribution to the long tradition in microeconomic theory research on endogenous product differentiation. Some models endogenise horizontal differentiation via location choice, be it over a line, a circle, or a spoke (Hotelling, 1929; D'Aspremont, Gabszewicz, and Thisse, 1979; Salop, 1979; Chen and Riordan, 2007). In others, quality choice is the measure instead (e.g. Shaked \& Sutton, 1982; Moorthy, 1988, Motta, 1993); a utility function uni-dimensional parameter that raises consumers' willingness to pay and (usually) marginal or fixed costs, and along which a firm locates optimally. An earlier strain of research on product variety as the optimal number of differentiated-product firms in a market is covered in some detail in Lancaster (1990). Models which combine both horizontal and vertical competition include von Ungern-Sternberg (1988) and Economides (1993).\\

% (1) where horizontal differentiation is defined as distance in the Salop (1979) circle, but firms may also endogenously select their own transport costs, understood as how close a good is to be perceived as "general purpose"; (2) which considers a model with one horizontal characteristic and one vertical characteristic with firms also distributed over a circle. The model in the present paper attempts to generalise these efforts, allowing for multi-dimensional choice on both the extensive and intensive side, for any number of goods and firms.\\

Lastly, this paper adds to a recent literature on games on networks. In this literature, it is closest to Ushchev and Zenou (2018) and Galeotti, Golub, and Goyal (2020). Ushchev and Zenou (2018) develops a model of price competition in a differentiated-product market in which products are linked through a network related to the Hessian matrix of the representative consumer's utility function, and shows that equilibrium prices depend on network structure. The present paper is in this sense a network formation game, as product design lays cross-price effect linkages between similar goods. Galeotti, Golub, and Goyal (2020) studies optimal interventions in network games with strategic spillovers, finding that these can be decomposed into orthogonal principal components determined by the network.

\section{Theoretical Framework}
\subsection{Setup}

In this paper, we focus on a simple version of the theoretical framework in Amir et al. (2017), with notation from Rodrigues (2025). A set of consumers with homogeneous preferences is modeled through a representative consumer (Gorman, 1959). The representative consumer's utility function is quasi-linear quadratic, as follows: 

\begin{equation}
    U(\boldsymbol{q},Y,\boldsymbol{p})=\boldsymbol{q}'\boldsymbol{\delta}-\frac{1}{2}\boldsymbol{q}'M\boldsymbol{q}-\phi (Y-\boldsymbol{q}'\boldsymbol{p})
\end{equation}

for $\boldsymbol{q}$ the representative consumer's consumption bundle; $\phi$ the representative consumer's price sensitivity, $\boldsymbol{\delta}$ the $N\times 1$ vector of initial marginal product utilities, $M$ the $N\times N$ positive definite Hessian matrix of the representative consumer's utility function, and $\boldsymbol{p}$ the $N\times 1$ price vector. $Y$ refers to the scalar income/wealth of the representative consumer.\\

Firms set prices in the second stage of a two-stage profit maximisation game. The optimal price, equilibrium quantity, and equilibrium level of profits are considered in the same two instances as before: where a set of $N$ single-product firms define prices individually, and where a single multi-product monopolist does so for its $N$ goods. For firm-level price-setting, the derivations of the FOCs (unadjusted for corner solutions for simplicity) yield the following equilibrium outcomes:

\begin{equation}\label{singotprice}
    \boldsymbol{p}^*=-\frac{1}{\phi}(\Omega+M^{-1})^{-1}M^{-1}\boldsymbol{\delta}\qquad
    \boldsymbol{q}(\boldsymbol{p}^*)=\Omega(\Omega+M^{-1})^{-1}M^{-1}\boldsymbol{\delta}
\qquad
    \boldsymbol{\pi}^*=\boldsymbol{p}^*\odot \boldsymbol{q}(\boldsymbol{p}^*)
\end{equation}

for $\Omega$ the $N\times N$ matrix representation of the diagonal of $M^{-1}$. For monopolist price-setting:

\begin{equation}\label{monopotprice}
    \boldsymbol{p}^*=-\frac{1}{2\phi}\boldsymbol{\delta}
\qquad
    \boldsymbol{q}(\boldsymbol{p}^*)=\frac{1}{2}M^{-1}\boldsymbol{\delta}
\qquad
    \Pi^*=-\frac{1}{4\phi}\boldsymbol{\delta}'M^{-1}\boldsymbol{\delta}
\end{equation}

For more details on the computations, the reader is encouraged to see Rodrigues (2026). In the first stage, firms determine endogenously what product characteristics to ascribe their goods. To understand product design as an optimisation problem for a profit-maximising firm, however, we first need to discuss how product characteristics enter the utility function of the representative consumer.\\

\subsection{Characteristics-Based Linear Demand}

Let initial marginal product utility be linear in product characteristics: $\boldsymbol{\delta}=X\boldsymbol{\beta}+\boldsymbol{v}$, for $X=[\boldsymbol{x}_1\ldots \boldsymbol{x}_K]$ a $N$ by $K$ matrix whose elements determine how much of a given observable characteristic $k$ good $n$ has; $\boldsymbol{\beta}$ a $K\times 1$ positive vector of the representative consumer's preference weights towards each of $K$ observable characteristics; and $\boldsymbol{v}$ a $N\times 1$ vector of independent, identically distributed, mean zero, unit variance, unobserved errors pertaining to demand shocks and latent product characteristics.\footnote{"Latent" refers both to variables unobservable to researchers and to variables outside the firm's control.} Assume $N\geq K$, i.e., there are (weakly) more goods than characteristics. \\

This is a commonplace functional form assumption for product utility. For product design purposes, it is however insufficiently detailed. The different columns of $X$ may be dependent on each other, making separate optimisation of each unrealistic. For example, we cannot separately choose over a good's mass without addressing how changes in mass impacts their volume; it may not be adequate to consider them as independent choice variables for a profit-maximising firm. If we wish to optimise over each separate characteristic, we must therefore re-state $X$ in such a way that each column isolates a independent lever, while mapping said levers to the original product's characteristics' physical (and/or technological or institutional) rigidities.\\
 
Via a QR decomposition\footnote{For this and other matrix algebra concepts used in this paper, see, e.g. Golub \& van Loan, 1983.} of $X$, we can write $\boldsymbol{\delta}=X\boldsymbol{\beta}+\boldsymbol{v}=ZR\boldsymbol{\beta}+\boldsymbol{v}=Z\boldsymbol{\tilde\beta}+\boldsymbol{v}$, for $Z\in\mathbb{R}^{N\times K}$ an orthonormal matrix, $R\in\mathbb{R}^{K\times K}$ an upper-triangular matrix, and $\boldsymbol{\tilde\beta}\in\mathbb{R}^{K\times 1}$ an orthogonally-corrected vector of preference weights towards each observable characteristic. Matrix $Z$ rewrites $X$ as orthonormal \textit{attributes}. Matrix $R$ on the other hand is a mechanical mapping; it describes how each attribute relates to each original characteristic. \\

To get an intuition, let us return to the point above about mass and volume. Mass and volume are related through the concept of density: $density=mass/volume$. Consider mass and volume vectors $\boldsymbol{m}$ and $\boldsymbol{v}$ (the latter not to be confused with the vector of demand shocks and latent product characteristics), for a range of goods of different densities $\rho$: $X=[\boldsymbol{v}\ \boldsymbol{m}]$:

\begin{equation}
\begin{aligned}
    X&=ZR\\
    &=\begin{bmatrix}\Large
        \frac{\boldsymbol{v}}{||\boldsymbol{v}||} & \Large\frac{\boldsymbol{m}-\overline{\rho}\boldsymbol{v}}{||\boldsymbol{m}-\overline{\rho}\boldsymbol{v}||}
    \end{bmatrix}\begin{bmatrix}
        ||\boldsymbol{v}|| & ||\boldsymbol{v}||\overline{\rho}\\0&||\boldsymbol{v}||\sigma_{\rho}
    \end{bmatrix}\\
    % &=\begin{bmatrix}\Large
    %     \frac{\boldsymbol{v}}{||\boldsymbol{v}||} & \Large\frac{(\boldsymbol{\rho}-\overline{\rho}\boldsymbol{1})\circ\boldsymbol{v}}{||(\boldsymbol{\rho}-\overline{\rho}\boldsymbol{1})\circ\boldsymbol{v}||}
    % \end{bmatrix}\begin{bmatrix}
    %     ||\boldsymbol{v}|| & ||\boldsymbol{v}||\overline{\rho}\\0&||\boldsymbol{v}||\sigma_{\rho}
    % \end{bmatrix}
\end{aligned}
\end{equation}

for $\overline{\rho} = \sum_{i=1}^Nw_i\rho_i$ the weighted mean density across all goods, $\sigma_{\rho}^2=\sum_{i=1}^nw_i(\rho_i-\overline{\rho})^2$ the weighted density variance, and $w_i=v_i^2/||\boldsymbol{v}||^2, \forall i$ the weights.\footnote{Volume $\boldsymbol{v}$ is squared in the weights because it enters $X$ directly and through $\boldsymbol{m}=\boldsymbol{\rho}\circ\boldsymbol{v}$.} In this way, $Z$ separately isolates two attributes for firms to decide on - product volume and deviation from (weighted) mean density, while $R$ codifies how mass of a given good depends on volume plus its own density heterogeneity.\\

There are a few different methods to orthonormalise vectors. One such method used above, Gram-Schmidt orthogonalisation, retains interpretability: attributes match their respective characteristics \textit{net} of the component explained by the preceding attribute(s). In our example, the first orthonormal direction isolates volume, while the second isolates mass net of its volume-proportional component, that is, net of the mass that would arise if all goods shared the common weighted-average density $\overline{\rho}$.\\

% One way to understand these attributes is follow how Gram-Schmidt orthogonalisations (one way to compute the QR decomposition) work. The first attribute $\boldsymbol{z}_1$ is a normalised version of $\boldsymbol{x}_1$; $\boldsymbol{z}_2$ is $\boldsymbol{x}_2$ corrected for its interdependence with $\boldsymbol{x}_1$: $\overline{\boldsymbol{z}}_2=\boldsymbol{x}_2-(\boldsymbol{z}_1'\boldsymbol{x}_2)\boldsymbol{z}_1$ (prior to normalisation to $\boldsymbol{z}_1$). The third attribute is obtained the same way: $\overline{\boldsymbol{z}}_3=\boldsymbol{x}_3-(\boldsymbol{z}_1'\boldsymbol{x}_3)\boldsymbol{z}_1-(\boldsymbol{z}_2'\boldsymbol{x}_3)\boldsymbol{z}_2$; and so on. The information embedded in multiple product characteristics is thus isolated onto a single attribute each time. In this example, e.g. $\boldsymbol{x}_3=\boldsymbol{Zr}_3$, for $\boldsymbol{r}_3= (||\overline{\boldsymbol{z}}_1||\boldsymbol{z}_1'\boldsymbol{x}_3;\ ||\overline{\boldsymbol{z}}_2||\boldsymbol{z}_2'\boldsymbol{x}_3;\ ||\overline{\boldsymbol{z}}_3||)$. \\

By restating $X$ such that each characteristic can be independently optimised through underlying attributes, we have made our linear demand model partially characteristics-based. However, endogenous product design not only affects the utility yielded by said products, but also the extent to which product compete. Products may be e.g. closer substitutes the more similar they are in their characteristics. From a utility function perspective, product similarity should play a role in determining how consumption of a given good satiates the consumer's need for said good and those sharing similar attributes. In linear demand, both satiation and substitution effects are driven by the latent Hessian of the utility function $U(\boldsymbol{q})$, represented by $M$.\\

What would $M$ look like if we assumed it to be also partially a function of observable product characteristics? Let $\hat M=E(M|X)$ and note the following:\\

\textit{\textbf{Lemma 1}: Any Hessian matrix that depends in part on observed product characteristics $X$ and imposes no arbitrary structure on unobserved dimensions admits the following representation:}

\begin{equation}
    \hat M = ZAZ'+\rho(I-ZZ')
\end{equation}

\textit{for $A$ the $K\times K$ reduced Hessian in characteristic coordinates, and $\rho$ a scalar governing the portion of $M$ unexplained by $Z$.}\footnote{Proofs of all Lemmas may be found in the Appendix.}\\

The matrix $A$ summarises how the observed characteristics contribute to the slope of demand and interact with one another. The scalar $\rho$ effectively weighs the relevance of a baseline-differentiation term.\\

We can go a little further. The $\hat M$ can be re-written as $\hat M=\rho I + Z\tilde DZ'$, for $\tilde D=D-\rho I$. As a symmetric matrix, we can write $\tilde D=U\Gamma U'$ via spectral decomposition. At last, we may write:

\begin{equation}
    \hat M = ZAZ'+\rho(I-ZZ')= (ZU)\Gamma(ZU)'+\rho I=  S\Gamma S'+\rho I= \sum_{k=1}^K\gamma_k\boldsymbol{s}_k\boldsymbol{s}_k'+\rho I
\end{equation}

for $U$ the orthogonal matrix whose columns are the eigenvectors of $\tilde D$, $\Gamma=\mathrm{diag}(\gamma_1,\dots,\gamma_K)$ the diagonal matrix of its eigenvalues, and $S=ZU$ the resulting orthonormal matrix of attribute directions in product space.\\

It is trivial to see that matrix $S=ZU$ is also an orthonormal matrix. In fact, matrix $S$ is just another orthonormalised representation for $X$: $X=ZR=SU^{-1}R$. Matrix $S$ is a re-statement of $Z$ in a new orthonormal basis of the same span - we therefore retain the \textit{attribute} naming. However, we must now keep in mind that an attribute is in fact a linear combination of different characteristics, and some attributes may load the same characteristics in opposite ways. There is no simple way to capture this.\\

Meanwhile, $\Gamma$ is a diagonal matrix whose elements pertain to the \textit{competitive salience} of each attribute; these are measures of the relative contribution of differentiation in a given attribute $\boldsymbol{s}_k$ to product positioning in the market, in the style of Bordalo, Gennaioli, and Shleifer (2013).\\

The matrix $\hat M$ is thus divided between a baseline idiosyncratic differentiation term $\rho I$ and a term measuring competition explainable from observable attributes firms can control.  Note that $M$ is positive definite if and only if $\rho +\gamma_K>0$. I assume $\gamma_k>0,\ \forall k\in\mathcal{K}$ to be the case throughout this paper - positive definiteness will then be automatic if at least one column exists. In general, $\gamma_p\neq\gamma_{q},\ \forall p\neq q=1,\ldots,K$. For simplicity, let $\rho=1$. To simplify notation, going forward we drop the hat in $\hat M$.\\

This way of formatting $M$ has multiple remarkable features that make it useful both to study the theoretical implications of assuming the relevance of product characteristics for competition, but also for empirical estimation. More on empirical estimation in the Appendix. Additional details on (i) how to obtain $S$ if $X$ is unobserved but we know $M$ and (ii) how $\hat M$ can be tied to $\mathbb{E}(\boldsymbol{\delta}\boldsymbol{\delta}')$ - i.e. the second raw moment of $\boldsymbol{\delta}$ - can also be found in the Appendix.\\

Before we continue, note:\\

\textit{\textbf{Lemma 2}: Product utility $\boldsymbol{\delta}=X\beta+\boldsymbol{v}=S\boldsymbol{b+\boldsymbol{v}}$, for $\boldsymbol{b}=U^{-1}R\beta>0$ the $K\times 1$ vector of attribute utilities.}\\

I incorporate also a quadratic cost per product characteristic within a good: $c_n(X)=\frac{1}{2}||\boldsymbol{x}_n||^2$. It can be shown that $c_n(\boldsymbol{x}_n)$ can be written in terms of $S$:

\begin{equation}
    c_n(\boldsymbol{x}_n)=\frac{1}{2}||\boldsymbol{x}_n||^2=\frac{1}{2}\boldsymbol{x}_n'\boldsymbol{x}_n=\frac{1}{2}\boldsymbol{s}_n'(U^{-1}RR'U)\boldsymbol{s}_n=\frac{1}{2}\boldsymbol{s}_n'C\boldsymbol{s}_n,\ \forall n
\end{equation}

for $C=U^{-1}RR'U$ a common term across goods. At last, we can state regarding expected profits $\mathbb{E}\pi$:\\

\textit{\textbf{Proposition 1}: For a full column rank $X$:}\\

\begin{equation}
    \max_{\{\boldsymbol{x}_{i}\}_{i\in \mathcal{G}_n}} \sum_{i\in \mathcal{G}_n}\mathbb{E}\pi_i(\boldsymbol{p}^*(X),X)=\max_{\{\boldsymbol{s}_{i}\}_{i\in \mathcal{G}_n}} \sum_{i\in \mathcal{G}_n}\mathbb{E}\pi_i(\boldsymbol{p}^*(S),S)
\end{equation}

\textit{Proof}: This result follows from $X=SU^{-1}R$, with $U^{-1}R$ kept fixed over the optimisation. $U$ is full rank by construction, as it reflects the eigenvectors of a positive definite matrix $A$, which itself follows from a positive definite $M$. The mapping from $X$ to $S$ is unique if and only if $U^{-1}R$ is full rank, which it will be if $R$ is full rank. $R$ is full rank if and only if $X$ is full column rank.\qedwhite\\

Full column rank is an undemanding assumption, requiring only that the observed characteristics not be exactly collinear across products. Put differently, it fails only when at least one characteristic adds no independent variation beyond the others.\footnote{Note that for $X$ to be full column rank, it is necessary (but not sufficient) that $K\leq N$, as assumed above.} Therefore, optimising product design via choice of characteristics $X$ can, under reasonable conditions, be said to be equivalent to doing so over orthonormalised attributes $S$.\\

Going forward, keep in mind the following definitions. Any attribute vector can be split into two, separately optimisable concepts: \textit{attribute intensity} and \textit{attribute orientation}. Attribute intensity pertains to the length of the attribute vector, e.g. $||\boldsymbol{s}_k||$; attribute orientation is the direction of the unit-normalised attribute vector: $||\boldsymbol{s}_k||^{-1}\boldsymbol{s}_k$. Furthermore, product differentiation will be defined in terms of \textit{observed} characteristics. Horizontal differentiation will be noted where, for e.g. a product pair, one of the goods has a relatively greater amount of a given attribute, and the other a relatively greater amount in another attribute. This can be best related to differences in attribute orientation. Vertical differentiation on the other hand will be noted where, for the same product pair, one good has (weakly) more of all attributes than the other. This is most observable in differences in attribute intensity. In both orderings, differentiation is qualitatively the same regardless of whether we define it in terms of attributes $\boldsymbol{s}$ or characteristics $\boldsymbol{x}$. The model otherwise assumes horizontal differentiation in unobserved characteristics, meaning that a good with weakly less of all attributes relative to all other goods can still be observed obtaining residual demand. 

\section{Optimal Product Design under Monopoly Pricing}

\subsection{One-Attribute Case}

The simplest case is that of a multi-product monopolist optimising over a single attribute ($K=1$) - we will use it to build intuition. We will prove the following:\\

\textit{\textbf{Proposition 2}: A multi-product product-designing monopolist optimising its products given a single attribute is indifferent in the extent of vertical differentiation across its goods, but never designs horizontal differentiation. Total attribute intensity across goods is increasing in attribute utility and decreasing in price sensitivity, the attribute cost coefficient, and attribute competitive salience.}\\

\textit{Proof}: For a single attribute, the Sherman-Morrison–Woodbury formula (Sherman and Morrison, 1949; Woodbury, 1950) simplifies:

\begin{equation}
    M^{-1}=I_N-\frac{\gamma\boldsymbol{s}\boldsymbol{s}'}{1+\gamma||\boldsymbol{s}||^2}
\end{equation}

Define $\boldsymbol{\delta}=\beta\boldsymbol{x}=(\beta||\boldsymbol{x}||)||\boldsymbol{x}||^{-1}\boldsymbol{x} =(\beta r)\boldsymbol{z =}b\boldsymbol{s}$. In this case, $\boldsymbol{s}=|\boldsymbol{z}|$ as $\boldsymbol{z}\tilde\beta=\boldsymbol{z}ub\Leftrightarrow b=\frac{1}{u}\tilde\beta$; $||u||=|u|=1$ necessarily as it must be orthonormal. Both $\boldsymbol{u}=\pm 1$ and $\boldsymbol{r}=||\boldsymbol{x}||$ are held fixed.\\

Note that:\\

\textit{\textbf{Lemma 3}: Expected monopoly profit $\mathbb{E}\Pi$ is independent of the idiossyncratic error $\boldsymbol{v}$}.\\

We therefore define $\boldsymbol{\delta}=b\boldsymbol{s}$ going forward WLOG. Then:

\begin{equation}
    \begin{aligned}
        \mathbb{E}\Pi &= (-\frac{1}{4\phi})\boldsymbol{\delta}'M^{-1}\boldsymbol{\delta}-\frac{1}{2}||\boldsymbol{x}||^2\\
        &=(-\frac{1}{4\phi}) b^2\boldsymbol{s}'(I_N-\frac{\gamma\boldsymbol{s}\boldsymbol{s}'}{1+\gamma||\boldsymbol{s}||^2})\boldsymbol{s}-\frac{1}{2}c||\boldsymbol{s}||^2\\
        &=(-\frac{1}{4\phi})\frac{ b^2||\boldsymbol{s}||^2}{1+\gamma||\boldsymbol{s}||^2}-\frac{1}{2}c||\boldsymbol{s}||^2
    \end{aligned}
\end{equation}

Notice how, in this context, $\mathbb{E}\Pi$ is shown to be dependent solely on the attribute intensity of $\boldsymbol{s}$. In other words, for a multi-product monopolist whose goods have a single characteristic, the distribution of said characteristic across the goods is not important. It is equally aggregate-profit-maximising for the monopolist to sell a single all-encompassing good $n$, a set of goods vertically differentiated in the set of attributes firms can control, or a range of fully-homogenous goods within the same set of attributes.\footnote{Equivalently and respectively: $\boldsymbol{s}^*=||\boldsymbol{s}||^*\boldsymbol{e}_n$; e.g., $\boldsymbol{s}^*=||\boldsymbol{s}||^*\sqrt{6/(N(N+1)(2N+1))}\cdot[N\ N-1\ \ldots\ 1]')$; $\boldsymbol{s}^*=||\boldsymbol{s}||^*/\sqrt{N}\cdot\boldsymbol{1}$.}\\

% To determine the optimal level of $||\boldsymbol{s}||$ to allocate the goods, we incorporate the quadratic attribute cost above. For a monopolist setting a single attribute, the relevant attribute cost is $\sum_{i=1}^N\frac{1}{2}c||\boldsymbol{x}||s_{n}^2=\frac{1}{2c}||\boldsymbol{x}||||\boldsymbol{s}||^2$

It can be shown that:

\begin{equation} ||\boldsymbol{s}||^*=\arg\max_{||\boldsymbol{s}||}\mathbb{E}\boldsymbol{\Pi}(\boldsymbol{p}^*,\boldsymbol{s})=
    \begin{cases}
    \frac{1}{\sqrt{\gamma}}\sqrt{-\frac{b^2}{2c\phi}-1}& \text{if }c<-b^2/2\phi\\
    \qquad\ \ \ \ 0 & \text{otherwise}
    \end{cases}
\end{equation}

We can then work backwards to see what this implies for $X$. For a single attribute, $\boldsymbol{x}=(\frac{\boldsymbol{x}}{||\boldsymbol{x}||})(||\boldsymbol{x}||)=\boldsymbol{z}r$. For $S$ to remain orthonormal, $|u|=1$. Therefore, $b=U^{-1}R\beta=r\beta$. Thus, we can find out what $\boldsymbol{x}^*$ looks like from the equality:
$\beta\boldsymbol{x}^*=b\boldsymbol{s}^*\Leftrightarrow\boldsymbol{x}^*=r\boldsymbol{s}^*$.\qedwhite\\

\subsection{Attribute Exclusivity}

In this section, I will prove the following:\\

\textit{\textbf{Proposition 3}: A multi-product monopoly optimising the design of its products given multiple exclusive attributes will load on these attributes in a manner proportional to their cost-weighted attribute utility. Attribute intensity across goods will be increasing in attribute utility and decreasing in price sensitivity, the attribute cost coefficient, and attribute competitive salience.}\\

Exclusivity means each attribute $K$ is set by (and loads on) exactly one good $n$:

\begin{equation}
\boldsymbol{s}_k = r_{nk}\boldsymbol{e}_{k}\qquad\Rightarrow\qquad\boldsymbol{s}_n=\boldsymbol{r}\cdot(\sum_{k\in\mathcal{A}_n}\boldsymbol{e}_k)=\boldsymbol{r}_n
\end{equation}

Define $\mathcal A_n$ as the set of all attributes exclusive to $n$. This allows the following simplification through the Sherman-Morrison–Woodbury formula:

\begin{equation}
M = I_N+\mathrm{diag}(\sum_{k\in\mathcal A_1}\gamma_k r_{1k}^2,\dots,\sum_{k\in\mathcal A_N}\gamma_k r_{Nk}^2)
\end{equation}

By construction, we have $\delta_n = \sum_{k\in\mathcal A_n} b_k r_{nk},\ \forall n$. Plugging the diagonal $M$ into the demand function and then calculating optimal prices gives, for each firm $n$:

\begin{equation} 
\begin{aligned}
&q_n^* = \frac{\delta_n}{2(1+\sum_{k\in\mathcal A_n}\gamma_k r_{nk}^2)}\\
&p_n^* = -\frac{1}{2\phi}\delta=-\frac{1+\sum_{k\in\mathcal A_n}\gamma_k r_{nk}^2}{\phi}q_n^*
\end{aligned}
\end{equation}

Product design over exclusive attributes separates each good into a separate, independent corner of the market. Hence the produce design problem decouples by good:

\begin{equation}
\begin{aligned}
&\max_{\{r_{nk}\}_{n,k}} \mathbb{E}\Pi\\
=&\max_{\{r_{nk}\}_{k\in\mathcal A_n}}\sum_{n=1}^N\mathbb{E}\pi_n(\boldsymbol{r}_n)\\
=&\sum_{n=1}^N\max_{\{r_{nk}\}_{k\in\mathcal A_n}}-\frac{1}{\phi}(1+\sum_{k\in\mathcal A_n}\gamma_k r_{nk}^2)\left(\frac{\sum_{k\in\mathcal A_n} b_k r_{nk}}{2(1+\sum_{k\in\mathcal A_n}\gamma_k r_{nk}^2)}\right)^{2}  -\frac{1}{2}\boldsymbol{r'_n}C\boldsymbol{r_n}\\
\end{aligned}
\end{equation}

As before, let $\hat r_k=\sqrt{\gamma_k}r_k$, $\hat b_k=\frac{1}{\sqrt{\gamma}}b_k$, and $\hat C=\Gamma^{-1/2}C\Gamma^{-1/2}$. Then, let $\boldsymbol{\hat r}_n=t_n\boldsymbol{d}_n$, for $t_n=||\boldsymbol{\hat r}_n||=||\boldsymbol{\gamma}^{\circ1/2}\boldsymbol{r}_n||$ and $||\boldsymbol{d}_n||=1$, these referring only to the space where $k\in\mathcal A_n,\ \forall n$.  We obtain:

\begin{equation}
    \begin{aligned}
        \mathbb{E}\pi_n =-\frac{1}{4\phi} \frac{(\boldsymbol{\hat b}'\boldsymbol{\hat r}_n)^2}{(1+\boldsymbol{\hat r}_n'\boldsymbol{\hat r}_n)}-\frac{1}{2} \boldsymbol{\hat r_n}'\hat C \boldsymbol{\hat r_n}=-\frac{1}{4\phi} \frac{t^2_n(\boldsymbol{\hat b}'\boldsymbol{d}_n)^2}{(1+t^2_n)}-\frac{1}{2}t^2_n\boldsymbol{d}_n'C\boldsymbol{d}_n
    \end{aligned}
\end{equation}

Let us write $\boldsymbol{\hat b}_n=\sum_{k\in\mathcal{A}_n}\boldsymbol{\hat b}\boldsymbol{e}_k$ WLOG: $\boldsymbol{\hat b}'\boldsymbol{\hat r}_n=\boldsymbol{\hat b}_n'\boldsymbol{\hat r}_n$ . Note then how, via the Cauchy-Schwarz inequality, $(\boldsymbol{\hat b_n}'\boldsymbol{d_n})^2\leq (\boldsymbol{\hat b_n}'C^{-1}\boldsymbol{\hat b_n})(\boldsymbol{d_n}'C\boldsymbol{d_n})$, holding with equality if $\boldsymbol{d}_n||\{C^{-1}\boldsymbol{\hat b}_n\}_{k\in\mathcal{A}_n}$. It thus follows that, in rehashing the same approach as in the previous section, the optimal $\boldsymbol{d}_n$ is as follows:

\begin{equation}
    \boldsymbol{s}^*_{n}=\boldsymbol{r}^*_{n}=t_n\cdot \Gamma^{-1/2}\boldsymbol{d}_{n}^*=t_n\cdot\Gamma^{-1/2}\frac{\{\hat C^{-1}\boldsymbol{\hat b}_n\}_{k\in\mathcal{A}_n}}{||\{\hat C^{-1}\boldsymbol{\hat b}_n\}_{k\in\mathcal{A}_n}||}=t_n\cdot\frac{\{ C^{-1}\boldsymbol{b}_n\}_{k\in\mathcal{A}_n}}{||\{\Gamma^{1/2}C^{-1}\boldsymbol{ b}_n\}_{k\in\mathcal{A}_n}||}
\end{equation}

What about $t_n$? As before, we can plug in $\boldsymbol{d}_n^*$ and perform the derivative of the resulting function in $t^2$, obtaining:

\begin{equation}
    -\frac{(1/4\phi) \boldsymbol{b}_n'C^{-1}\boldsymbol{b}_n}{(1+t_n^2)^2}=\frac{1}{2}\Leftrightarrow t^{2*}_n=\max\Bigg\{\sqrt{-\frac{1}{2\phi}\boldsymbol{b}_n'C^{-1}\boldsymbol{b}_n}-1,0\Bigg\}
\end{equation}

Note that $t^*\geq0$ by construction, but $\boldsymbol{s}_n$ need not be element-by-element positive. This concludes our proof.\qedwhite\\

\subsection{Attribute Non-Exclusivity}

Let us now consider the multi-attribute case under non-exclusivity. We will prove the following:\\

\textit{\textbf{Proposition 4}: A multi-product product-designing monopolist optimising its products given multiple attributes does not pursue horizontal differentiation. It is indifferent to the number of goods it produces or the extent of their vertical differentiation; attribute intensity across goods is proportional to their cost-weighted attribute utility; increasing in attribute utility and decreasing in price sensitivity, the attribute cost coefficient, and attribute competitive salience.}\\

\textit{Proof}: Consider the pair of attribute vectors $\boldsymbol{s}_i$ and $\boldsymbol{s}_j$, describing how much of attributes $i$ and $j$ each good has. By the Cauchy-Schwarz inequality and orthonormality of the initial representation, $\boldsymbol{s}_i'\boldsymbol{s}_j\in[-1,1]$, with equality only attainable if $\boldsymbol{s}_i||\boldsymbol{s}_j$. To analyse how a change in $\boldsymbol{s}_i'\boldsymbol{s}_j,\ \forall i,j$ affects expected profits in our model without affecting any other attribute pair (including those which relate to either of these attributes), we consider shifts that allow $\boldsymbol{s}_i$ and $\boldsymbol{s}_j$ to remain orthogonal to all other pairs. We can do this by varying the pair along the 2-D plane that is orthogonal to the span of the remaining vectors, setting, e.g.:

\begin{equation}
    \boldsymbol{s}_i=\boldsymbol{e}_i\qquad\boldsymbol{s}_j=\cos\theta\boldsymbol{e}_j+\sin\theta\boldsymbol{e}_i,\qquad\theta\in[-\frac{\pi}{2},\frac{\pi}{2}]
\end{equation}

This can be done WLOG. We can organise any pair of $\boldsymbol{s}$ in this form as (i) their multiplication remains the same - we have only rearranged what part of $\boldsymbol{s}_i'\boldsymbol{s}_j$ stays on which side, in the same way that e.g. $12=2\times6=3\times4$ - and (ii) it does not affect any other attribute pair.\\

The first point can be generalised by noting that any $\boldsymbol{s}_i'\boldsymbol{s}_j$ can be re-written as a function of the maximum value it may take ($1$ in this case since we are keeping attribute intensity across all columns of $S$ fixed) and a parameter $\theta$:

\begin{equation}
    \boldsymbol{s}_i'\boldsymbol{s}_j=\sin\theta\in[-1,1]
\end{equation}

The split into $\boldsymbol{s}_i$ and $\boldsymbol{s}_j$, with $\boldsymbol{e}_i$ and $\boldsymbol{e}_j$ follows.\\

To see the second point: for any $k\neq i,j$, assume we start with $\boldsymbol{s}_k\perp\boldsymbol{e}_i$ and $\boldsymbol{s}_k\perp\boldsymbol{e}_j$, such that all $\boldsymbol{s}$ are representationally orthogonal to each other prior to optimisation. Since $\boldsymbol{s}_i(\theta)$ and $\boldsymbol{s}_j(\theta)$ lies in the plane spanned by $\boldsymbol{e}_i$ and $\boldsymbol{e}_j$, we have, $\forall i,j\neq k=1,\ldots,K$ and $\theta$:

\begin{equation}
    \boldsymbol{s}_k'\boldsymbol{s}_i(\theta)=\boldsymbol{s}_k'\boldsymbol{e}_i=0\qquad\boldsymbol{s}_k'\boldsymbol{s}_j(\theta)=\cos\theta \boldsymbol{s}_k'\boldsymbol{e}_j+\sin\theta \boldsymbol{s}_k'\boldsymbol{e}_i=0
\end{equation}

The proof is trivial: if $\boldsymbol{s}_k'\boldsymbol{e}_i=0$, true by construction, then (i) $\boldsymbol{s}_k'\boldsymbol{s}_i(\theta)=0$; (ii) $\sin\theta \boldsymbol{s}_k'\boldsymbol{e}_i=0$; and (iii) for $\boldsymbol{s}_k'\boldsymbol{s}_j(\theta)$ to be $0$ for one $\theta$ (such that orthogonality holds at some representation of $S$), then $\cos\theta \boldsymbol{s}_k'\boldsymbol{e}_j=0$, which must be true for any $\theta$.\\

Consider the Sherman–Morrison–Woodbury formula for matrices:

\begin{equation}\label{shermanmorr}
    M^{-1}=(I_N+\sum_{k=1}^K\gamma_k\boldsymbol{s}_k\boldsymbol{s}_k')^{-1}=(I+S\Gamma S')^{-1}=I-S(\Gamma^{-1}+S'S)^{-1}S'
\end{equation}

In respect of partial differentiation, we can thus allow $\boldsymbol{s}_i'\boldsymbol{s}_j$ to vary, while every other pair stays at zero. This means that $\Gamma^{-1}+S'S$ remains block diagonal in what is relevant for optimisation. If we place $i$ and $j$ as the first columns $S$:

\begin{equation}
    \Gamma^{-1}+S'S = \begin{bmatrix}
        A & 0\\ 0 & B
    \end{bmatrix},
    \qquad 
    A = \begin{bmatrix}
        \frac{1}{\gamma_i}+1& \boldsymbol{s}_i'\boldsymbol{s}_j\\ \boldsymbol{s}_i'\boldsymbol{s}_j & \frac{1}{\gamma_j}+1
    \end{bmatrix}
    \qquad
    B = \begin{bmatrix}
        \frac{1}{\gamma_{3}}+1& 0 &\cdots& 0\\
        0 &0 & \cdots& 0\\
        \vdots & \vdots&\ddots & \vdots\\
        0 &0&\cdots& \frac{1}{\gamma_{N}}+1
    \end{bmatrix}
\end{equation}

This means:

\begin{equation}\label{Ainv}
    (\Gamma^{-1}+S'S)^{-1}=
    \begin{bmatrix} 
        A^{-1}& 0\\ 0& B^{-1}
    \end{bmatrix}
    \qquad\ \text{for}\ \qquad A^{-1} = \frac{1}{\Delta}
    \begin{bmatrix}
        \frac{1}{\gamma_j}+1& -\boldsymbol{s}_i'\boldsymbol{s}_j\\ -\boldsymbol{s}_i'\boldsymbol{s}_j& \frac{1}{\gamma_i}+1
    \end{bmatrix}
\end{equation}

for $\Delta=(1/\gamma_i+1)(1/\gamma_j+1)-(\boldsymbol{s}_i'\boldsymbol{s}_j)^2$. Since we have that $\boldsymbol{s}_i'\boldsymbol{s}_j\in[-1,1]$, $\Delta$ is strictly positive. From here, note that, from (\ref{monopotprice}):

\begin{equation}\label{mechahint}
\begin{aligned}
    \mathbb{E}\Pi &= (-\frac{1}{4\phi})\boldsymbol{\delta}'M^{-1}\boldsymbol{\delta}-\frac{1}{2}\sum_{n=1}^N||\boldsymbol{x}_n||^2\\
    &=(-\frac{1}{4\phi})\big[||\boldsymbol{\delta}||^2-\boldsymbol{\delta}'S(\Gamma^{-1}+S'S)^{-1}S'\boldsymbol{\delta}\big]-\frac{1}{2}\sum_{n=1}^N\boldsymbol{s_n}'C\boldsymbol{s_n}\\
    &=(-\frac{1}{4\phi})\big[\boldsymbol{
b}'S'S\boldsymbol{b}-\boldsymbol{
b}'S'S(\Gamma^{-1}+S'S)^{-1}S'S\boldsymbol{b}\big]-\frac{1}{2}\sum_{n=1}^N\boldsymbol{s_n}'C\boldsymbol{s_n}
\end{aligned}
\end{equation}

reflecting the fact that, with the exception of the cost term, $\mathbb{E}\boldsymbol{\Pi}$ depends only on attribute intensities ($||\boldsymbol{s}_i||, \ \forall i=1,\ldots K$) and angles (how attribute orientation differs relative to others, rather than in absolute: $\boldsymbol{s}_i'\boldsymbol{s}_j, \ \forall i,j=1,\ldots K$).\\

Regarding the cost term, we can express it in terms of attribute vectors (rather than per-good vectors) as follows:

\begin{equation}
    \frac{1}{2}\sum_{n=1}^N\boldsymbol{s}_n'C_n\boldsymbol{s}_n=\frac{1}{2}\sum_{n=1}^N\sum_{k=1}^K\sum_{l=1}^K\{C_n\}_{kl}s_{nk}\boldsymbol{s}_nl=\frac{1}{2}\sum_{k=1}^K\sum_{l=1}^K\boldsymbol{s}_k'D_{kl}\boldsymbol{s}_l
\end{equation}

with $D_{kl},\forall k,l=1,\ldots K$ a diagonal matrix which places the $C_{n,kl}$ element in the $n$-th diagonal. The expression can be further simplified for a symmetric and identical $C$ across all goods. Partialling out all terms in the cost term which depend on attributes $i$ and $j$ and accounting for orthogonality, we get:

\begin{equation}
    \bigg\{\sum_{n=1}^Nc_n(X)\bigg\}_{i,j}=\frac{1}{2}C_{ii}||\boldsymbol{s}_i||^2+\frac{1}{2}C_{jj}||\boldsymbol{s}_j||^2+C_{ij}\boldsymbol{s}_i'\boldsymbol{s}_j
\end{equation}

We can then partial out the attribute vectors' contribution to $\mathbb{E}\Pi$:

\begin{equation}\label{pairwisepi}\small
    \begin{aligned}
        \mathbb{E}\Pi_{\{i,j\}}
        =
        &\frac{
        b_i^2\frac{1}{\gamma_i}\left[\frac{1}{\gamma_j}+1-(\boldsymbol{s}_i'\boldsymbol{s}_j)^2\right]
        +
        b_j^2\frac{1}{\gamma_j}\left[\frac{1}{\gamma_i}+1-(\boldsymbol{s}_i'\boldsymbol{s}_j)^2\right]
        +
        2b_ib_j\frac{1}{\gamma_i}\frac{1}{\gamma_j}\boldsymbol{s}_i'\boldsymbol{s}_j
        }{-4\phi\Delta}-\bigg\{\sum_{n=1}^Nc_n(X)\bigg\}_{i,j}
    \end{aligned}
\end{equation}
\normalsize

% Differentiating this expression relative to $\theta$:

% \begin{equation}
%     \begin{aligned}
%         &\frac{\partial\mathbb{E}\Pi_{\{i,j\}}}{\partial \theta}=\\&\frac{2\sin\theta[b_i^2\gamma_j+b_j^2\gamma_i]+2b_ib_j(1/\gamma_i+1/\gamma_j+1)[(1/\gamma_i+1)(1/\gamma_j+1)-\sin^2\theta]}{-4\phi\Delta^2}-C_{ij}\cos \theta
%     \end{aligned}
% \end{equation}

Differentiating this expression relative to $\theta$:

\begin{equation}\small
    \begin{aligned}
        &\frac{\partial\mathbb{E}\Pi_{\{i,j\}}}{\partial \theta}=\\&\frac{\Bigg[
        2b_ib_j\frac{1}{\gamma_i}\frac{1}{\gamma_j}
        \left[
        \left(\frac{1}{\gamma_i}+1\right)\left(\frac{1}{\gamma_j}+1\right)
        +\sin^2\theta
        \right]
        -2\sin\theta\left[
        b_i^2\left(\frac{1}{\gamma_i}\right)^2\left(\frac{1}{\gamma_j}+1\right)
        +
        b_j^2\left(\frac{1}{\gamma_j}\right)^2\left(\frac{1}{\gamma_i}+1\right)
        \right]
        \Bigg]\cos\theta
        }{-4\phi\Delta^2}\\&
        -C_{ij}\cos\theta
    \end{aligned}
\end{equation}

By definition, $\cos\theta\geq0$. Therefore, we observe stationary points at $\cos\theta=0$, equivalent to the $\boldsymbol{s}_i'\boldsymbol{s}_j=\pm1$ cases. In other words, full or exact opposite orientation alignment are candidate optima. However, note how, for a sufficiently small $C$, $\mathbb{E}\Pi_{\{i,j\}}$ is greater at every positive value of $\theta$ than at the corresponding negative value; $\frac{\partial\mathbb{E}\Pi_{\{i,j\}}}{\partial \theta}$ is also greater at every negative $\theta$ than at every positive; and at $\theta=0$, $\frac{\partial\mathbb{E}\Pi_{\{i,j\}}}{\partial \theta}>0$. For a sufficiently small $C$, this rules out global optima in the region where $\boldsymbol{s}_i'\boldsymbol{s}_j<0$. \\ 

This result by itself does not rule out $\boldsymbol{s}_i'\boldsymbol{s}_j<0$ or an interior global solution, i.e. some degree of horizontal differentiation. We explore this further in Section 3.6., when we introduce asymmetric costs. There, we prove, from a different perspective, that for the multi-product monopolist $\boldsymbol{s}_1||\ldots||\boldsymbol{s}_N$ with positive proportionality. If this is true, then there are no interior solutions in the attribute vectors: $\boldsymbol{s}_i'\boldsymbol{s}_j=\pm1$ at most.\\

Assume this to be the case here.\\

This leaves two remaining questions: (i) what is the optimal orientation that all columns of $S$ should be parallel to; and (ii) what is the optimal attribute intensity. Regarding the first of these, based on the parallelism of the $\boldsymbol{s}$ vectors, allow every $\boldsymbol{s}_k=r_k\boldsymbol{y}$, for $r_k$ a scalar and $\boldsymbol{y}>0$ the common factor, $||\boldsymbol{y}||=1$. This setup effectively reduces the multi-factor case:

\begin{equation}
    \boldsymbol{\delta}=\sum_{k=1}^Kb_ks_k=\sum_{k=1}^Kb_kr_k\boldsymbol{y}=(\boldsymbol{b}'\boldsymbol{r})\boldsymbol{y}
\end{equation}

\begin{equation}
    M^{-1}= I-\frac{\sum_{k=1}^K(\gamma_kr_k)^2}{1+\sum_{k=1}^K(\gamma_kr_k^2)}\boldsymbol{y}\boldsymbol{y}'
\end{equation}

so that: 

\begin{equation}
    \begin{aligned}
        \mathbb{E}\Pi &=(-\frac{1}{4\phi}) \frac{\sum_{k=1}^K(b_kr_k)^2}{1+\sum_{k=1}^K(\gamma_kr_k^2)}-\frac{1}{2}\sum_{i=1}^N\boldsymbol{s_n'}C\boldsymbol{s_n}\\
        &=(-\frac{1}{4\phi}) \frac{(\boldsymbol{b}'\boldsymbol{r})^2}{1+\boldsymbol{r}'\Gamma\boldsymbol{r}}-\frac{1}{2}||\boldsymbol{y}||^2\boldsymbol{r'}C\boldsymbol{r}\\
        &=(-\frac{1}{4\phi}) \frac{(\boldsymbol{\hat b}'\boldsymbol{\hat r})^2}{1+\boldsymbol{\hat r}'\boldsymbol{\hat r}}-\frac{1}{2}\boldsymbol{\hat r}'\hat C\boldsymbol{\hat r}
    \end{aligned}
\end{equation}

for $\hat r_k=\sqrt{\gamma_k}r_k$ and $\hat b_k=\frac{1}{\sqrt{\gamma_k}}b_k,\ \forall k=1,\ldots,K$. We also use $\hat C=\Gamma^{-1/2}C\Gamma^{-1/2},\ \forall k$. Thus, we have, via a further decomposition, $\boldsymbol{\hat r}=t\boldsymbol{d}$, with $t=||\boldsymbol{\hat r}||=||\boldsymbol{\gamma}^{\circ1/2}\boldsymbol{r}||>0$ setting intensity, and $\boldsymbol{d}$ setting sign and orientation, with $||\boldsymbol{d}||=1$:

\begin{equation}
    \begin{aligned}
        \mathbb{E}\Pi =(-\frac{1}{4\phi}) \frac{(\boldsymbol{\hat b}'\boldsymbol{\hat r})^2}{1+\boldsymbol{\hat r}'\boldsymbol{\hat r}}-\frac{1}{2}\boldsymbol{\hat r}'\hat C\boldsymbol{\hat r}=(-\frac{1}{4\phi}) \frac{t^2(\boldsymbol{\hat b}'\boldsymbol{d})^2}{1+t^2}-\frac{1}{2}t^2\boldsymbol{d}'\hat C\boldsymbol{d}
    \end{aligned}
\end{equation}

The only relevant choice variable is $r_k,\ \forall k$, the norm of each vector. What is the optimal $\boldsymbol{\hat r}$? Via Cauchy-Schwarz, $(\boldsymbol{\hat b}'\boldsymbol{d})^2\leq (\boldsymbol{\hat b}'\hat C^{-1}\boldsymbol{\hat b})(\boldsymbol{d}'\hat C\boldsymbol{d})$, holding with equality if $\boldsymbol{d}\parallel \hat C^{-1}\boldsymbol{\hat b}$. This is key, as we can show that it maximises $\mathbb{E}\Pi$:

\begin{equation}
\begin{aligned}
    \mathbb{E}\Pi&\leq(-\frac{1}{4\phi}) \frac{t^2(\boldsymbol{\hat b}'\hat C^{-1}\boldsymbol{\hat b})(\boldsymbol{d}'\hat C\boldsymbol{d})}{1+t^2}-\frac{1}{2}t^2\boldsymbol{d}'\hat C\boldsymbol{d}\\
    &=\bigg((-\frac{1}{4\phi}) \frac{t^2(\boldsymbol{\hat b}'\hat C^{-1}\boldsymbol{\hat b})}{1+t^2}-\frac{1}{2}t^2\bigg)\boldsymbol{d}'\hat C\boldsymbol{d}\\
    &\propto\bigg((-\frac{1}{4\phi}) \frac{t^2(\boldsymbol{\hat b}'\hat C^{-1}\boldsymbol{\hat b})}{1+t^2}-\frac{1}{2}t^2\bigg)\boldsymbol{\hat b}\hat C^{-1}\boldsymbol{\hat b}
\end{aligned}
\end{equation}

which is independent of $\boldsymbol{d}$, such that the expression has a zero gradient in that variable. In other words, for any fixed $t$:

\begin{equation}
    \boldsymbol{r}^*=t\cdot\Gamma^{-1/2}\boldsymbol{d}^*=t\cdot\Gamma^{-1/2}\frac{\hat C^{-1}\boldsymbol{\hat b}}{||\hat C^{-1}\boldsymbol{\hat b}||}=t\cdot\frac{C^{-1}\boldsymbol{ b}}{|| \Gamma^{1/2}C^{-1}\boldsymbol{ b}||}
\end{equation}

$C$ need not be diagonally dominant, hence the sign uncertainty for $\boldsymbol{r}^*$. What about $t$? Working from the above, we can plug in $\boldsymbol{d}^*$ and perform the derivative of the resulting (concave) function in $t^2$, obtaining:

\begin{equation}
    -\frac{(1/4\phi) \boldsymbol{b}'C^{-1}\boldsymbol{b}}{(1+t^2)^2}=\frac{1}{2}\Leftrightarrow t^{2*}=\max\Bigg\{\sqrt{-\frac{1}{2\phi}\boldsymbol{b}'C^{-1}\boldsymbol{b}}-1,0\Bigg\}
\end{equation}

We reach a result equivalent to the one-attribute case - $\mathbb{E}\boldsymbol{\Pi}$ is independent of $\boldsymbol{y}$. We can conclude that, where full or inverse alignment is optimal, for this setting too is it equally aggregate-profit-maximising for the monopolist to sell a single all-encompassing good $n$, a vertically-differentiated set of goods in the set of observed attributes, or a range of homogenous goods in the set of observable attributes. For as long as $||\boldsymbol{y}||=1$, aggregate profit is maximised independently by every $\boldsymbol{r}$. \qedwhite \\

\subsection{Discussion}

What is the intuition for our results? Equation (\ref{mechahint}) provides a hint. There are three mechanisms driving effects of product design on expected profits. First, there is an implicit complementarity in how attributes are stacked. The norm of the initial product utility vector is highest when goods are parallel. For e.g. the two-attribute case:

\begin{equation}
    ||\boldsymbol{\delta}||^2=\boldsymbol{b}'S'S\boldsymbol{b}=b_1(b_1+\boldsymbol{s}_i'\boldsymbol{s}_jb_2)+b_2(\boldsymbol{s}_i'\boldsymbol{s}_jb_1+b_2)\qquad\frac{\partial ||\boldsymbol{\delta}||^2}{\partial \boldsymbol{s}_i'\boldsymbol{s}_j}=2b_1b_2>0
\end{equation}

This also drives indifference in the number of goods, as gains in utility through one good or the other are equivalent.\\

The second mechanism is competitive pressure. Cross-price effects are highest at $\boldsymbol{s}_i'\boldsymbol{s}_j=\pm1$ and attributes are (anti-)parallel (see (\ref{Ainv})). As can be seen in (\ref{mechahint}), if $\Gamma^{-1}=0$, the two effects would cancel out exactly ($(S'S)^{-1}S'S=I$). With $\Gamma^{-1}>0$, however, this effect never compensates the first mechanism.\\

Further alignment has a net positive result on optimal demand - this follows almost immediately from a well-behaved (strictly concave) utility function. Alignment stops the monopolist from spreading attractiveness across different directions and instead lets it push the whole product line towards the direction the consumer likes most. The whole line is pulled toward the same attractive part of product space.\\

The third mechanism is that of cost $c_n$. In some cases, elements of $C$ can be negative. This happens when the two attributes act as cost-substitutes: producing more of both together is cheaper than one might infer from adding their standalone costs only, because some of the underlying characteristic adjustments overlap or undo each other. When it happens, this incentivises further alignment.

\section{Optimal Product Design under Single-Product Firm Pricing}

Now consider $N$ single-product firms, one good per firm, each with up to $K$ attributes. Each firm sets its own price. As before, we work with expected profits and define $\boldsymbol{\delta}=S\boldsymbol{b}$ for simplicity. This is WLOG for the optimal attribute orientation proof below, and qualitatively unimportant for our discussion of optimal attribute intensity.

% This setting is far more complex than that stated above; in what follows, I there consider two extreme equilibria: the case where (i) all firms choose to optimise only over exclusive attributes (asymmetric equilibrium); and (ii) where attributes are non-exclusive but the all firms choose the same attributes (symmetric equilibrium). Lastly, I simulate numerical solutions for $S$ and study the resulting equilibria; I show that all equilibria found this way match the symmetric equilibrium in attribute orientation across goods, and differ only in attribute intensity per good, confirming our general finding that firms optimally pursue vertical differentiation in controllable product attributes.

\subsection{Attribute Exclusivity}

We start with multi-attribute product design, but with each attribute being exclusively assigned to only one product. This can be akin to assuming heterogeneity in the product characteristics each firm is allowed control over. From our earlier example with mass and volume, the first firm could be given control over only the volume of their good (first attribute), picking over different standardised materials/formulations; whereas the second firm could set how far to deviate from mean density (second attribute), perhaps through a fixed format which they could then vary in formulation or material composition. In other words, we could define attribute exclusivity such that one is only allowed to make the package bigger at average density, while the other can only change density/formulation holding package size roughly fixed.\\

I will prove the following:\\

\textit{\textbf{Proposition 5}: A single-product firm optimising the design of its product given multiple exclusive attributes will load on these attributes in a manner proportional to their cost-weighted attribute utility. Attribute intensity across goods will be increasing in attribute utility and decreasing in price sensitivity, the attribute cost coefficient, and attribute competitive salience.}\\

\textit{Proof}: The proof mirrors that of the multi-product monopolist setting. Exclusivity means each attribute $K$ is set by (and loads on) exactly one firm $n$:

\begin{equation}
\boldsymbol{s}_k = r_{nk}\boldsymbol{e}_{k}\qquad\Rightarrow\qquad\boldsymbol{s}_n=\boldsymbol{r}\cdot(\sum_{k\in\mathcal{A}_n}\boldsymbol{e}_k)=\boldsymbol{r}_n
\end{equation}

Define $\mathcal A_n$ as the set of all attributes exclusive to $n$. This allows the following simplification:

\begin{equation}
M = I_N+\mathrm{diag}(\sum_{k\in\mathcal A_1}\gamma_k r_{1k}^2,\dots,\sum_{k\in\mathcal A_N}\gamma_k r_{Nk}^2)
\end{equation}

By construction, we have $\delta_n = \sum_{k\in\mathcal A_n} b_k r_{nk},\ \forall n$. Plugging the diagonal $M$ into the demand function and then calculating optimal prices gives, for each firm $n$:

\begin{equation} 
\begin{aligned}
&q_n^* = \frac{\delta_n}{2(1+\sum_{k\in\mathcal A_n}\gamma_k r_{nk}^2)}\\
&p_n^* = -\frac{1}{2\phi}\delta=-\frac{1+\sum_{k\in\mathcal A_n}\gamma_k r_{nk}^2}{\phi}q_n^*
\end{aligned}
\end{equation}

This is effectively the monopoly setting. Product design over exclusive attributes separates each good into a separate, independent corner of the market. Hence the produce design problem decouples by firm:

\begin{equation}
\max_{\{r_{nk}\}_{k\in\mathcal A_n}} \mathbb{E}\pi_n=\max_{\{r_{nk}\}_{k\in\mathcal A_n}}-\frac{1}{\phi}(1+\sum_{k\in\mathcal A_n}\gamma_k r_{nk}^2)\left(\frac{\sum_{k\in\mathcal A_n} b_k r_{nk}}{2(1+\sum_{k\in\mathcal A_n}\gamma_k r_{nk}^2)}\right)^{2}  -\frac{1}{2}\boldsymbol{r'_n}C\boldsymbol{r_n}
\end{equation}

with quadratic attribute costs defined as follows:

\begin{equation}
    \frac{1}{2}\boldsymbol{s}_n'C\boldsymbol{s}_n=\frac{1}{2}\boldsymbol{r'_n}C\boldsymbol{r_n}
\end{equation}

As before, let $\hat r_k=\sqrt{\gamma_k}r_k$, $\hat b_k=\frac{1}{\sqrt{\gamma}}b_k$, and $\hat C=\Gamma^{-1/2}C\Gamma^{-1/2}$. Then, let $\boldsymbol{\hat r}_n=t_n\boldsymbol{d}_n$, for $t_n=||\boldsymbol{\hat r}_n||=||\boldsymbol{\gamma}^{\circ1/2}\boldsymbol{r}_n||$ and $||\boldsymbol{d}_n||=1$, these referring only to the space where $k\in\mathcal A_n,\ \forall n$.  We obtain:

\begin{equation}
    \begin{aligned}
        \mathbb{E}\pi_n =-\frac{1}{4\phi} \frac{(\boldsymbol{\hat b}'\boldsymbol{\hat r}_n)^2}{(1+\boldsymbol{\hat r}_n'\boldsymbol{\hat r}_n)}-\frac{1}{2} \boldsymbol{\hat r_n}'\hat C \boldsymbol{\hat r_n}=-\frac{1}{4\phi} \frac{t^2_n(\boldsymbol{\hat b}'\boldsymbol{d}_n)^2}{(1+t^2_n)}-\frac{1}{2}t^2_n\boldsymbol{d}_n'C\boldsymbol{d}_n
    \end{aligned}
\end{equation}

Let us write $\boldsymbol{\hat b}_n=\sum_{k\in\mathcal{A}_n}\boldsymbol{\hat b}\cdot\boldsymbol{e}_k$ WLOG: $\boldsymbol{\hat b}'\boldsymbol{\hat r}_n=\boldsymbol{\hat b}_n'\boldsymbol{\hat r}_n$ . Note then how, once again with Cauchy-Schwarz, $(\boldsymbol{\hat b_n}'\boldsymbol{d_n})^2\leq (\boldsymbol{\hat b_n}'C^{-1}\boldsymbol{\hat b_n})(\boldsymbol{d_n}'C\boldsymbol{d_n})$, holding with equality if $\boldsymbol{d}_n||\{C^{-1}\boldsymbol{\hat b}_n\}_{k\in\mathcal{A}_n}$. It thus follows that, in rehashing the same approach as in the previous section, the optimal $\boldsymbol{d}_n$ is as follows:

\begin{equation}
    \boldsymbol{s}^*_{n}=\boldsymbol{r}^*_{n}=t_n\cdot \Gamma^{-1/2}\boldsymbol{d}_{n}^*=t_n\cdot\Gamma^{-1/2}\frac{\{\hat C^{-1}\boldsymbol{\hat b}_n\}_{k\in\mathcal{A}_n}}{||\{\hat C^{-1}\boldsymbol{\hat b}_n\}_{k\in\mathcal{A}_n}||}=t_n\cdot\frac{\{ C^{-1}\boldsymbol{b}_n\}_{k\in\mathcal{A}_n}}{||\{\Gamma^{1/2}C^{-1}\boldsymbol{ b}_n\}_{k\in\mathcal{A}_n}||}
\end{equation}

What about $t_n$? As before, we can plug in $\boldsymbol{d}_n^*$ and perform the derivative of the resulting function in $t^2$, obtaining:

\begin{equation}
    -\frac{(1/4\phi) \boldsymbol{b}_n'C^{-1}\boldsymbol{b}_n}{(1+t_n^2)^2}=\frac{1}{2}\Leftrightarrow t^{2*}_n=\max\Bigg\{\sqrt{-\frac{1}{2\phi}\boldsymbol{b}_n'C^{-1}\boldsymbol{b}_n}-1,0\Bigg\}
\end{equation}

Once again, $t^*\geq0$ by construction, but $\boldsymbol{s}_n$ need not be element-by-element positive. This concludes our proof.\qedwhite\\

\subsection{Attribute Non-Exclusivity}

Let us now turn to a more complicated setting: multi-attribute design without exclusivity. With non-exclusivity, any firm may load any attribute $K$.\\

We will prove the following:\\

\textit{\textbf{Proposition 6}: Competition between single-product price-setting firms optimising the design of their products given multiple attributes reach a unique symmetric equilibrium. While their starting point matches that of the same setting in the monopoly pricing case - i.e. an optimal attribute vector proportional to their cost-weighted attribute utility - competing firms tilt their attribute vector across greater attribute dimensions. Competition makes firms more balanced in the attributes they load, particularly by emphasizing the attributes with lesser competitive salience over those with lower weighted costs as the number of firms increases.}\\

\textit{Proof}: Many of the simplifications that have been made possible by either product exclusivity or joint ownership are no longer available to us. Instead, we take advantage of the symmetry of the firms. Assume a symmetric candidate equilibrium in which all firms choose the same attribute vector:

\begin{equation}
    \boldsymbol{s}_n=\boldsymbol{r}=t\cdot\Gamma^{-1/2}\boldsymbol{d}\in\mathbb{R}^K,\ \forall n
\end{equation}

for $||\boldsymbol{d}||=1$ and $t=||\boldsymbol{\hat r}||\geq0$ as before. First, I will let this be true while allowing a given firm $n$ to deviate from $\boldsymbol{r}$ by setting $\boldsymbol{s}_n=\overline{t}\cdot\Gamma^{-1/2}\boldsymbol{d}$. Define for this purpose $\boldsymbol{t}=\sum_{i\neq n}\boldsymbol{e}_i+\overline{t}\boldsymbol{e}_n$. The goal will be determining the optimal symmetric $t^*$ via $\frac{\partial\mathbb{E}\pi_n}{\partial \overline{t}}|_{\overline{t}=t}=0$ Second, I will allow a given firm $n$ to deviate by setting $\boldsymbol{\overline{d}}_n$ and verify the conditions under which $\frac{\partial\mathbb{E}\pi_n}{\partial \boldsymbol{\overline{d}}}|_{\boldsymbol{\overline{d}}=\boldsymbol{d}}=0$.\\

Note:

\begin{equation}
    M=I+S\Gamma S'=I+\hat S \hat S'=I+(\boldsymbol{t}\boldsymbol{ d}') (\boldsymbol{t}\boldsymbol{d}')'=I+\boldsymbol{t}\boldsymbol{d}'\boldsymbol{d}\boldsymbol{t}'=I+\boldsymbol{t}\boldsymbol{t}'
\end{equation}

We can go further. By the Sherman–Morrison–Woodbury formula for the 1-factor case:

\begin{equation}
    M^{-1}=I-\frac{\boldsymbol{t}\boldsymbol{t}'}{1+\overline{t}^2+(N-1)t^2}
\end{equation}

with $\Omega_{nn}=\frac{1+(N-1)t^2}{1+\overline{t}^2+(N-1)t^2}$ and $\Omega_{-n,-n}=\frac{1+\overline{t}^2+(N-2)t^2}{1+\overline{t}^2+(N-1)t^2}$. Then, for $\boldsymbol{\delta}=\boldsymbol{t\boldsymbol{d'}\boldsymbol{\hat b}}=\boldsymbol{\hat b}'\boldsymbol{d}\boldsymbol{t}$ and:

\begin{equation}\small\label{m1delta}
    M^{-1}\boldsymbol{\delta}=\bigg(I-\frac{\boldsymbol{t}\boldsymbol{t}'}{1+\overline{t}^2+(N-1)t^2}\bigg)\boldsymbol{\hat b}'\boldsymbol{d}\boldsymbol{t}=\boldsymbol{\hat b}'\boldsymbol{d}\bigg(I-\frac{\boldsymbol{t}\boldsymbol{t}'}{1+\overline{t}^2+(N-1)t^2}\bigg)\boldsymbol{t}=\frac{1}{1+\overline{t}^2+(N-1)t^2}\boldsymbol{\hat b}'\boldsymbol{d}\boldsymbol{t}
\end{equation}

we may re-write the optimal price equation as follows, separately defining $\boldsymbol{p}^*=(p_n^*\quad \boldsymbol{p}_{-n}^*)'$, for $\boldsymbol{p}_{-n}^*=p_{-n}^*\boldsymbol{1}$, due to symmetry:

\begin{equation}
\begin{aligned}
    &\boldsymbol{p}^*=-\frac{1}{\phi}(\Omega+M^{-1})^{-1}M^{-1}\boldsymbol{\delta}\\
    \Leftrightarrow&(\Omega+M^{-1})\boldsymbol{p}^*=-\frac{1}{\phi}M^{-1}\boldsymbol{\delta}\\
    \Leftrightarrow&
    \begin{pmatrix}
        2(1+(N-1)t^2)p_n^*-(N-1)\overline{t}tp_{-n}^*\\
        -\overline{t}tp_n^*+(2+2\overline{t}^2+(N-2)t^2)p_{-n}^*
    \end{pmatrix}
    =
    \begin{pmatrix}
        -\frac{1}{\phi}\boldsymbol{\hat b}'\boldsymbol{d}\overline{t}\\
        -\frac{1}{\phi}\boldsymbol{\hat b}'\boldsymbol{d}t\\
    \end{pmatrix}
\end{aligned}
\end{equation}

Solving the system of equations, optimal prices are:

\begin{equation}
    p_n^*= -\frac{1}{\phi}\boldsymbol{\hat b}'\boldsymbol{d}\overline{t}(\frac{2+2\overline{t}^2+(2N-3)t^2}{\Delta})
\end{equation}
\begin{equation}
    p_{-n}^*=-\frac{1}{\phi}\boldsymbol{\hat b}'\boldsymbol{d}t(\frac{\overline{t}^2+2(1+(N-1)t^2)}{\Delta})
\end{equation}

for $\Delta=2(1+(N-1)t^2)(2+2\overline{t}^2+(N-2)t^2)-(N-1)\overline{t}^2t^2$. Optimal quantities follow from $\boldsymbol{q}(\boldsymbol{p}^*)=-\phi\Omega\boldsymbol{p}^*$:

\begin{equation}
    q_n(\boldsymbol{p}^*)=-\phi\Omega_{nn} p_n^*\qquad q_{-n}(\boldsymbol{p}^*)=-\phi\Omega_{-n,-n} p_{-n}^*
\end{equation}

and expected firm profit will be:

\begin{equation}
\begin{aligned}
    \mathbb{E}\pi_n&=\ p_n^*q_n(\boldsymbol{p}^*)-\frac{1}{2}\boldsymbol{\hat s}_n'\hat C\boldsymbol{\hat s}_n\\
    &=-\phi\Omega_{nn} p_n^{*2}-\frac{1}{2}\overline{t}^2\boldsymbol{d}'\hat C\boldsymbol{d}\\
    &=-\frac{1}{\phi}(\frac{1+(N-1)t^2}{1+\overline{t}^2+(N-1)t^2})(\frac{2+2\overline{t}^2+(2N-3)t^2}{\Delta})^2(\boldsymbol{\hat b}'\boldsymbol{d}\overline{t})^2-\frac{1}{2}\overline{t}^2\boldsymbol{d}'\hat C\boldsymbol{d}
\end{aligned}
\end{equation}

Then, taking the FOC of $\mathbb{E}\pi_n$ on $\overline{t}$:

\begin{equation}\label{tought}
\begin{aligned}
    \frac{\partial\mathbb{E}\pi_n}{\partial \overline t}\Bigg|_{\overline{t}=t}&=-\frac{1}{\phi}(\boldsymbol{\hat b}'\boldsymbol{d})^2t\frac{2(1+(N-1)t^2)P_N(t)}{(1+Nt^2)^2(2+(N-1)t^2)^3(2+(2N-1)t^2)}-t\boldsymbol{d}'\hat C\boldsymbol{d}\\
    &=t(F_N(t)(\boldsymbol{\hat b}'\boldsymbol{d})^2-\boldsymbol{d}'\hat C\boldsymbol{d})
\end{aligned}
\end{equation}

for $P_N(t)=4+2(5N-4)t^2+(N-1)(8N-7)t^4+(N-1)(2N^2-5N+1)t^6$. It can be shown that, for $N\geq2$, if $F_N(t)>0$ then $\frac{\partial F_N(t)}{\partial t}<0$, as both depend on if $P_N(t)\geq0$ (signs flip otherwise). Any interior symmetric solution must satisfy $\frac{\partial\mathbb{E}\pi_n}{\partial t}\Big|_{\overline{t}=t}=0$ and $P_N(t)\geq0$ (as $\boldsymbol{d}'\hat C\boldsymbol{d}>0$\footnote{This is because $C$ is positive definite, since $C = U'X'XU$ and $X$ is full column rank.}). Since on that region we have just stated that the first term on the RHS of (\ref{tought}) is strictly decreasing in $t$, the equation $\frac{\partial\mathbb{E}\pi_n}{\partial t}\Big|_{\overline{t}=t}=0$ must have at most one positive solution. We can furthermore show that, for $t^*>0$ and evaluating at $t=0$, we need:

\begin{equation}\label{secondcond}
\begin{aligned}
   &(\boldsymbol{\hat b}'\boldsymbol{d})^2F_N(0)\geq\boldsymbol{d}'\hat C\boldsymbol{d}\\
   \Leftrightarrow&-\frac{1}{2\phi}(\boldsymbol{\hat b}'\boldsymbol{d})^2\geq\boldsymbol{d}'\hat C\boldsymbol{d}\\
   \Rightarrow&-\frac{1}{2\phi}(\boldsymbol{\hat b}'\hat C^{-1}\boldsymbol{\hat b})(\boldsymbol{d}'\hat C\boldsymbol{d})\geq \boldsymbol{d}'\hat C\boldsymbol{d}\\
   \Rightarrow&-\frac{1}{2\phi}\boldsymbol{ b}' C^{-1}\boldsymbol{ b}\geq1
\end{aligned}
\end{equation}

a condition aligned with the other setups we have considered. Even without a closed-form solution, we have been able to argue for the existence of a unique symmetric optimal attribute intensity.\footnote{Were we to restore $\boldsymbol{v}$ to $\boldsymbol{\delta}$, we would require: $(\boldsymbol{\hat b}'\boldsymbol{d})^2\geq(3N+1)/(2N+2)\xrightarrow{\infty}1.5$ so that the first term of (\ref{tought}) is decreasing in $t$; and the condition in place of (\ref{secondcond}) be $-\frac{1}{2\phi}((\boldsymbol{\hat b}'\boldsymbol{d})^2-1)\geq\boldsymbol{d}'\hat C\boldsymbol{d}$.}\\

We now move on to proving that there is a unique symmetric equilibrium attribute orientation. To achieve this, we stick with $\boldsymbol{s}_{-n}=t\cdot\Gamma^{-1/2}\boldsymbol{d}$, and now set, for a firm $n$, $\boldsymbol{s}=t\boldsymbol{g}$. Vector $\boldsymbol{g}$ will have a unique structure:

\begin{equation}
    \boldsymbol{g}=\cos(\theta) \boldsymbol{d}+\sin(\theta) \boldsymbol{h},\quad \boldsymbol{h}\perp \boldsymbol{d},\quad ||\boldsymbol{h}||=1
\end{equation}

such that the deviation we consider is only in orientation. Unlike a format used earlier in the paper to analyse attribute angles, this approach allows $\theta$ to stay within the range $[0,\pi]$ ($0$ if $\boldsymbol{g}'\boldsymbol{d}=1$, $\pi$ if $\boldsymbol{g}'\boldsymbol{d}=-1$), facilitating interpretation of derivatives relative to this parameter. We can then write $M$ as follows:

\begin{equation}
    M(\theta)=\begin{bmatrix}
        1+t^2 & t^2(g(\theta)'\boldsymbol{d})\boldsymbol{1}_{N-1}'\\
        t^2(g(\theta)'\boldsymbol{d})\boldsymbol{1}_{N-1} & I_{N-1}+t^2\boldsymbol{1}_{N-1}\boldsymbol{1}_{N-1}'
    \end{bmatrix}=\begin{bmatrix}
        1+t^2 & t^2\cos(\theta)\boldsymbol{1}_{N-1}'\\
        t^2\cos(\theta)\boldsymbol{1}_{N-1} & I_{N-1}+t^2\boldsymbol{1}_{N-1}\boldsymbol{1}_{N-1}'
    \end{bmatrix}
\end{equation}

We can then show that, for small deviations of $\boldsymbol{g}$ from $\boldsymbol{d}$, the impact on $M$ is second-order:

\begin{equation}
    M(\theta)-M(0)=t^2(g(\theta)'\boldsymbol{d}-1)\begin{bmatrix}
        0 & \boldsymbol{1}'_{N-1}\\
        \boldsymbol{1}_{N-1} & 0
    \end{bmatrix}=t^2(\cos(\theta)-1)\begin{bmatrix}
        0 & \boldsymbol{1}'_{N-1}\\
        \boldsymbol{1}_{N-1} & 0
    \end{bmatrix}
\end{equation}

As $\frac{\partial (M(\theta)-M(0))}{\partial \theta}\big|_{\theta=0}\propto-sin(0)=0$, the effect of a firm deviating from $\boldsymbol{d}$ on $M$, and therefore $M^{-1}$, $\Omega$, and $(\Omega+M^{-1})^{-1}$, is negligible. Consider now $\boldsymbol{\delta}$. Its components are:

\begin{equation}
    \delta_n(\theta)=t\boldsymbol{\hat b}\boldsymbol{g}(\theta)\quad\delta_{-n}=t\boldsymbol{\hat b}\boldsymbol{d}
\end{equation}

The effect of a changing attribute orientation on $\boldsymbol{\delta}$ is first-order:

\begin{equation}
    \frac{\partial \boldsymbol{\delta}}{\partial \theta}\Bigg|_{\theta=0}=t(\boldsymbol{\hat b}'h)\boldsymbol{e}_n
\end{equation}

Bringing these results together:

\begin{equation}
    \frac{\partial M^{-1}(\theta)\boldsymbol{\delta}(\theta)}{\partial \theta}\Bigg|_{\theta=0}=t(\boldsymbol{\hat b}'h)(\boldsymbol{e}_n-\frac{t^2}{1+Nt^2}\boldsymbol{1})
\end{equation}

The second term in brackets is recognisable from (\ref{m1delta}), though the $t$ are now identical across firms. From before, we also have the following optimal pricing equation and derivative:

\begin{equation}
    p_n(0)=-\frac{t}{\phi[2+(N-1)t^2]}(\boldsymbol{\hat b}'\boldsymbol{d})
\end{equation}

\begin{equation}
    \frac{\partial p^*_n(\theta)}{\partial \theta}\Bigg|_{\theta=0}=-\frac{t(2+(3N-2)t^2+(N-1)^2t^4)}{\phi (2+(N-1)t^2)(2+(2N-1)t^2)}(\boldsymbol{\hat b}'\boldsymbol{h})\\
%     \Leftrightarrow&(\Omega+M^{-1})\boldsymbol{p}^*=-\frac{1}{\phi}M^{-1}\boldsymbol{\delta}
\end{equation}

and optimal quantity equation and derivative:

\begin{equation}
    q_n(\boldsymbol{p}^*,\theta)=-\phi\Omega_{nn}(\theta)p_n(\theta)\qquad \frac{\partial q_n(\boldsymbol{p}^*,\theta)}{\partial\theta}\Bigg|_{\theta=0}=-\phi\bigg(\frac{1+(N-1)t^2}{1+Nt^2}\bigg) \frac{\partial p^*_n(\theta)}{\partial \theta}\Bigg|_{\theta=0}
\end{equation}

Lastly, note:

\begin{equation}
    \frac{\partial }{\partial \theta}\frac{1}{2}\boldsymbol{g}_n(\theta)' \hat C\boldsymbol{ g}_n(\theta)\Bigg|_{\theta=0}=t^2\boldsymbol{h}'\hat C\boldsymbol{d}
\end{equation}

These are useful for computing the impact of changing attribute orientation by one firm on its own expected profits:

\begin{equation}
\begin{aligned}
    \frac{\partial \mathbb{E}\pi_n}{\partial \theta}\Bigg|_{\theta=0}&=t\frac{\partial g(\theta)}{\partial \theta}\Bigg|_{\theta=0}\bigg(p_n(0)\frac{q_n(\boldsymbol{p}^*,\theta)}{\partial\theta}\Bigg|_{\theta=0}+q_n(\boldsymbol{p}^*,0)\frac{\partial p_n^*(\theta)}{\partial\theta}\Bigg|_{\theta=0}-t\hat C\boldsymbol{d}\bigg)\\
    &=t\boldsymbol{h}'\bigg(-2\phi\bigg(\frac{1+(N-1)t^2}{1+Nt^2}\bigg)p_n(0)\frac{\partial p_n^*(\theta)}{\partial\theta}\Bigg|_{\theta=0}-t\hat C\boldsymbol{d}\bigg)\\
    &=-t^2\frac{2(1+(N-1)t^2)(2+(3N-2)t^2+(N-1)^2t^4)}{\phi(1+Nt^2)(2+(N-1)t^2)^2(2+(2N-1)t^2)}(\boldsymbol{\hat b}'\boldsymbol{d})(\boldsymbol{\hat b}'\boldsymbol{h})-t^2\boldsymbol{h}'\hat C\boldsymbol{d}\\
    &=t^2(K_n(\boldsymbol{\hat b}'\boldsymbol{d})(\boldsymbol{\hat b}'\boldsymbol{h})-\boldsymbol{h}'\hat C\boldsymbol{d})
\end{aligned}
\end{equation}

for $K_N=-\frac{2(1+(N-1)t^2)(2+(3N-2)t^2+(N-1)^2t^4)}{\phi(1+Nt^2)(2+(N-1)t^2)^2(2+(2N-1)t^2)}$. Thus, the symmetric equilibrium orientation $\boldsymbol{d}$ will be a stationary point if and only if:

\begin{equation}
    \boldsymbol{h}'(K_N(\boldsymbol{\hat b}'\boldsymbol{d})\boldsymbol{\hat b}-\hat C\boldsymbol{d})=0\qquad \boldsymbol{h}\perp\boldsymbol{d}
\end{equation}

This is equivalent to the existence of some scalar $\mu$ such that:

\begin{equation}
    \mu \boldsymbol{d}=K_N(\boldsymbol{\hat b}'\boldsymbol{d})\boldsymbol{\hat b}-\hat C \boldsymbol{d}
\end{equation}

which we can re-arrange by multiplying both sides by $\boldsymbol{d}'$ and applying the optimisation condition for $t^*$:

\begin{equation}
\begin{aligned}
    &\ \mu \boldsymbol{d}=K_N(\boldsymbol{\hat b}'\boldsymbol{d})\boldsymbol{\hat b}-\hat C \boldsymbol{d}\\
    \Leftrightarrow&\ \mu=K_N(\boldsymbol{\hat b}'\boldsymbol{d})^2-\boldsymbol{d}'\hat C\boldsymbol{d}\\
    \Leftrightarrow&\ \mu=(K_N-F_N)(\boldsymbol{\hat b}'\boldsymbol{d})^2
\end{aligned}
\end{equation}

It can be shown that, for $N\geq 2$, $K_N>F_N>0$ for all $t>0$, meaning that, for the purpose of this setting, $\mu>0$. We can re-arrange the initial equation to find the optimal orientation:

\begin{equation}
\begin{aligned}
    &\ \mu \boldsymbol{d}=K_N(\boldsymbol{\hat b}'\boldsymbol{d})\boldsymbol{\hat b}-\hat C \boldsymbol{d}\\
    \Leftrightarrow&\ (\hat C+\mu I) \boldsymbol{d}=K_N(\boldsymbol{\hat b}'\boldsymbol{d})\boldsymbol{\hat b}\\
    \Leftrightarrow&\ \boldsymbol{d}^*=K_N(\boldsymbol{\hat b}'\boldsymbol{d})(\hat C+\mu I)^{-1}\boldsymbol{\hat b}\\
    \Rightarrow&\ 
    \boldsymbol{d}^*\ ||\ (\hat C+\mu I)^{-1}\boldsymbol{\hat b}\\
    \Rightarrow&\ 
    \boldsymbol{d}^*\ ||\ \Gamma^{1/2} (C+\mu \Gamma)^{-1}\boldsymbol{b}\\
\end{aligned}
\end{equation}

If $N=1$, as in the monopoly case, $\mu=0$, yielding that original result. While we have not proved a global optimum, we have found a unique symmetric stationary equilibrium candidate; there is at most one interior symmetric equilibrium.\\

It may appear that this symmetric equilibrium only implies that firm's orientations are parallel, not that they are positively proportional. However, two goods cannot coexist with negatively proportional $\boldsymbol{d}$; if so, for at least one $\delta_n<0$. The orientation rule above therefore pins down positive proportionality. This concludes the proof.\qedwhite\\

\subsection{Discussion}

The intuition behind the attribute exclusivity result is relatively straightforward. If firms optimise only over exclusive attributes, we have a diagonal $M$ - goods are independent and there are no cross-price effects. Since each firm's design problem only affects product utility and demand slope, firms distribute their products attributes symmetrically following the same orientation rule as we saw in the monopoly case - though applied to their own exclusive attributes. This is a farily unique finding: two firms, in the same product category, with similar characteristics, ultimately act like monopolists - their strategic design levers are independent, and they make independent decisions. When a firm changes its exclusive attribute, it does not create a design-mediated cross-price effect on rivals. The mechanism can best be understood in terms of commitment, e.g., firms committing to product design optimisation in a way that minimises inter-comparability. This way, firms stop being close substitutes on the margins they can actually move.\\

Attribute non-exclusivity has its own unique implications. Because the inverted term will be relatively larger than the original $\hat C^{-1}$ for the lower-cost directions, this appears to favour greater dispersion of attributes per good than e.g. the multi-product monopoly and the attribute-exclusive single-product firms settings. Compared with a monopolist (in product or attributes), competitive firms tilt their attribute vector in all directions and spread themselves across greater attribute dimensions, proportionally to their salience. Competition makes firms more balanced in the attributes they load. The mechanism behind this is as follows: $\mu$ acts as a uniform penalty on concentrating too hard in the otherwise most favorable directions. In effect, it mirrors a greater competitive pressure, which pushes firms away in an attempt to minimise it along the most cost-effective popular attributes. Note how $K_N-F_N$ increases and then plateaus as the number of firms $N$ increases - suggesting that greater competition biases consumers increasingly away from the attributes for which their costs are lower and towards those with lesser competitive salience.\\

For the attribute orientation case, we have only identified $\boldsymbol{s}^*=\boldsymbol{t}^*\boldsymbol{d}^*$ as a stationary point. Going beyond this is impractical and generally intractable. Via numerical simulation, we can however obtain a greater degree of confidence that this is indeed the unique symmetric equilibrium - and in fact is the only fixed point as a result of a convergence in iterated best-responses from different stating points. The graph below reflects such a numerical simulation, whereby firms converge to the symmetric equilibrium we have identified.\\

\begin{figure}[H]
    \centering
    \caption{Selected sample attribute choice simulations under symmetric costs}
    %\caption{Per-supermarket transaction trends}
    \includegraphics[scale=0.3]{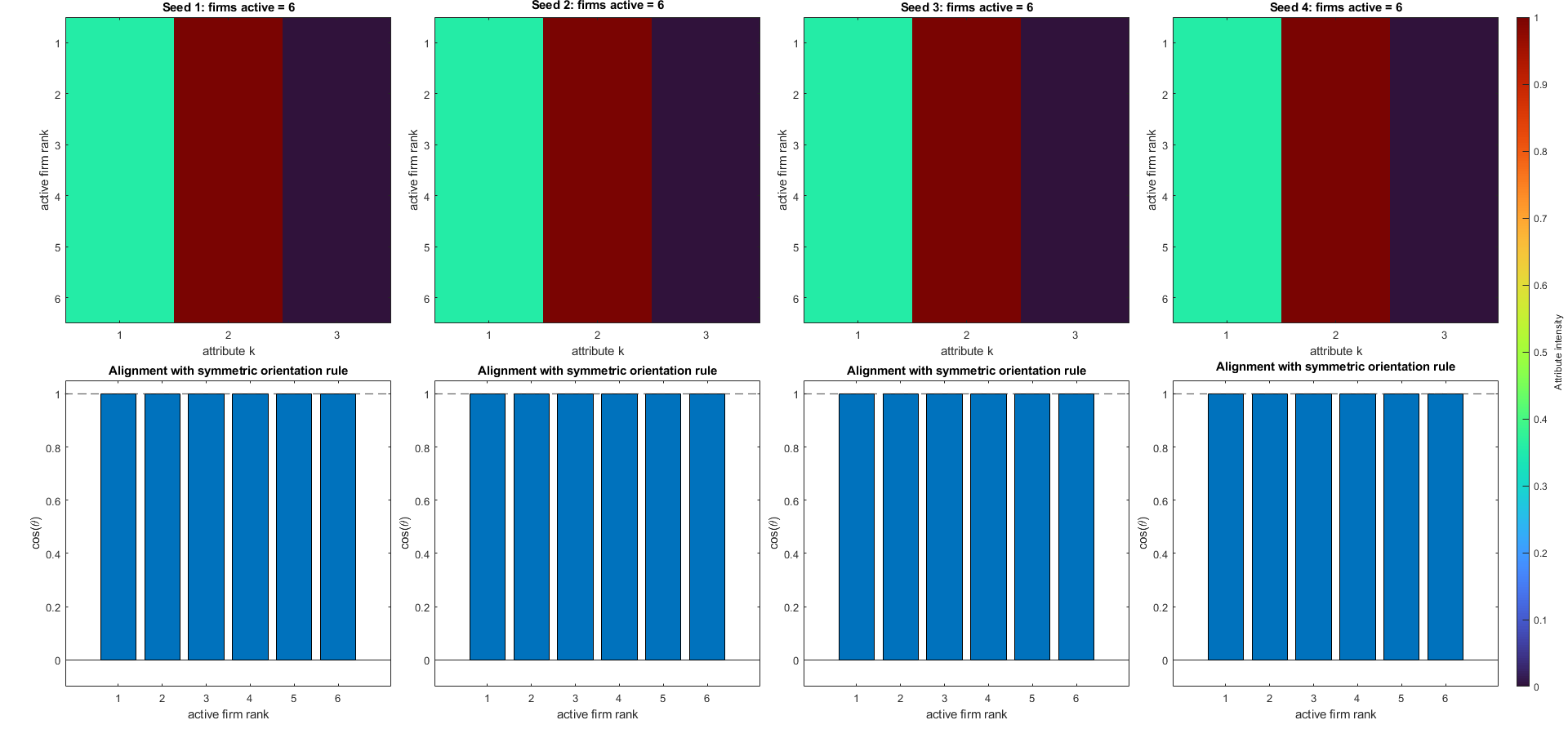}\\\RaggedRight\singlespacing
\footnotesize Notes: Heatmap colours defines attribute intensity; alignment measured via cosine between attribute orientation and proposed orientation rule. Simulations converged after 113, 123, 131, and 130 iterations. More details on the simulation in the Appendix.\\
\end{figure}
\normalsize

This doubles down on our finding that firms, even where faced with competitive pressure, still behave symmetrically, such that horizontal differentiation does not arise. Yet this may be because we have introduced no source of firm-/product-level heterogeneity. We consider this in the next section.

\section{Firm- and Product-level Heterogeneity in Attribute Costs}

There are many sources of competitive asymmetry, for example: asymmetry in  market power (as in Stackelberg competition), information (e.g. Akerlof, 1970),  consumer captivity (e.g. Varian, 1980), capacity constraints (e.g. Kreps and Scheinkman, 1983), marginal costs (e.g. Blume, 2003), and most recently in data (Rhodes and Zhou, 2024) and network effects (Peitz and Sato, 2025).\\

So far, we have analysed a general case where firms decide upon different attributes with different costs. Nonetheless, we have so far excluded varying attribute costs per good/firm. The more complex version of this is a setting where:

\begin{equation}
    \frac{1}{2}\boldsymbol{x}_n'\Sigma_n\boldsymbol{x}_n=\frac{1}{2}\boldsymbol{s}_n'(U^{-1}R\Sigma_nR'U)\boldsymbol{s}_n=\frac{1}{2}\boldsymbol{s}_n'C_n\boldsymbol{s}_n
\end{equation}

This induces some interesting outcomes to both single-product firms and multi-product monopolists. Let us consider each in turn. For single-product firms, I will show the following:\\

\textit{\textbf{Proposition 7}: Under firm-level heterogeneity in attribute costs, single-product firms optimising the design of their products given multiple attributes pursue horizontal differentiation, adjusting their designs both to weigh cost-weighted attribute utility and the strategic design of their competitors.}\\

\textit{Proof}: One simple way to see how this affects single-product firms is to consider the FOC for a given firm $n$ in the single-product firms setting under attribute non-exclusivity:

\begin{equation}\label{parallelsingprod}
    C_n\boldsymbol{s}_n=\frac{\partial \mathbb{E}\pi_n^*}{\partial\delta_n}\boldsymbol{b}+2
\frac{\partial \mathbb{E}\pi_n^*}{\partial M_{nn}}\Gamma\boldsymbol{s}_n+2
\sum_{i\neq n}^N\frac{\partial \mathbb{E}\pi_n^*}{\partial M_{ni}}\Gamma\boldsymbol{s}_i
\end{equation}

This completes the proof.\qedwhite\\

Note how, as $M\rightarrow I$, the second and third terms, describing a strategic incentives towards/away from differentiation, go away. We can then write, for $\alpha_n> 0$ (since, by (\ref{singotprice}) and $M\rightarrow I$, ${\partial \mathbb{E}\pi_n^*}/{\partial\delta_n}>0$):

\begin{equation}
    C_n\boldsymbol{s}_n=\alpha_n\boldsymbol{b}\qquad\Leftrightarrow\qquad C_1\boldsymbol{s}_1\ ||\ \ldots\ ||\ C_N\boldsymbol{s}_N\quad\&\quad \boldsymbol{s}_n||C_n^{-1}\boldsymbol{b},\quad \forall n=1,\ldots, N
\end{equation}

Numerical assessments confirm a tendency towards this adjusted orientation rule:

\begin{figure}[H]
    \caption{Selected sample attribute choice simulations under asymmetric costs}
    \centering
    %\caption{Per-supermarket transaction trends}
    \includegraphics[scale=0.3]{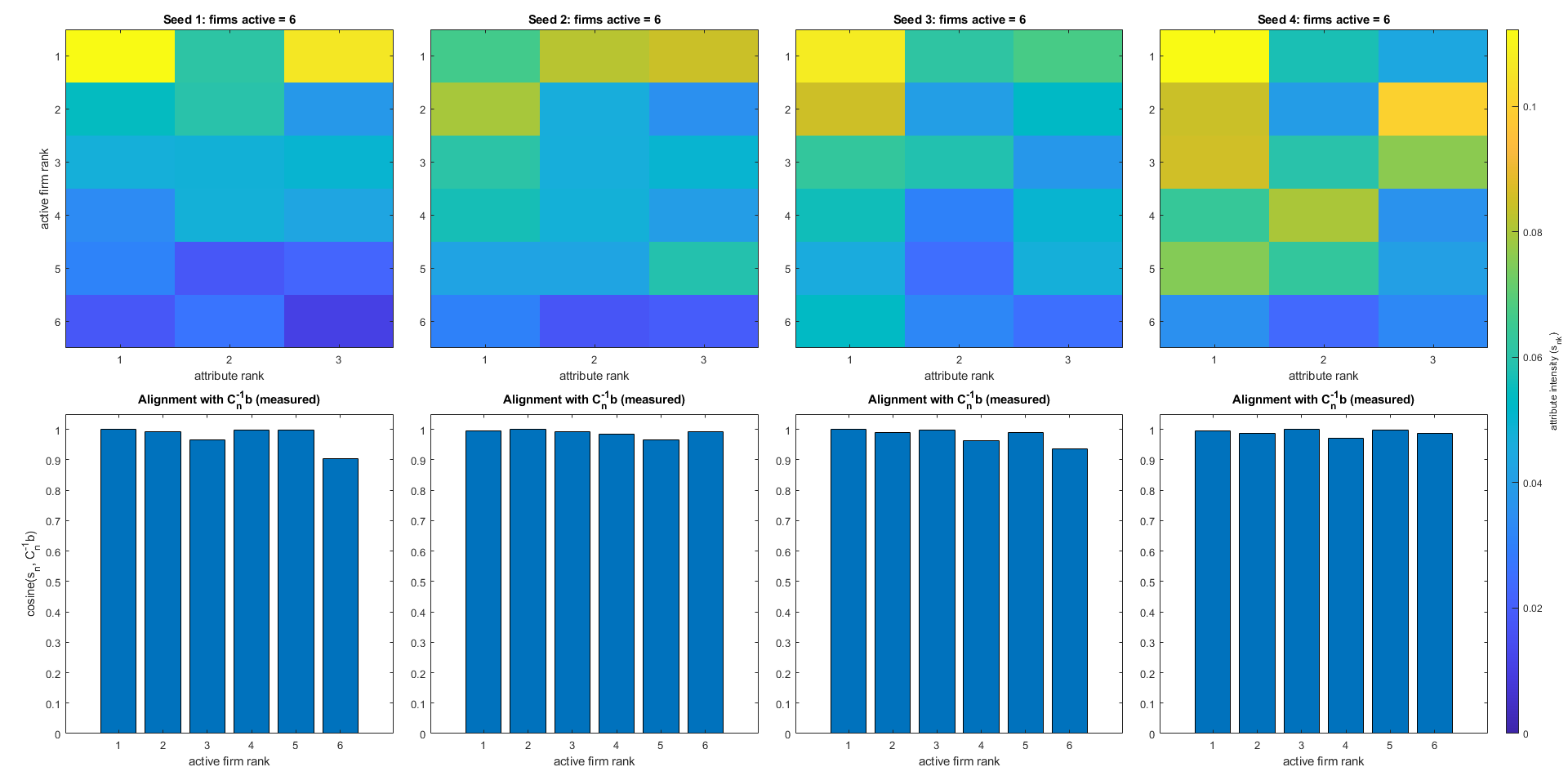}\\\RaggedRight\singlespacing
\footnotesize Notes: Heatmap colours defines attribute intensity; alignment measured via cosine between attribute orientation and proposed orientation rule. Simulations converged after 826, 341, 1121, and 321 iterations. More details on the simulation in the Appendix.\\
\end{figure}
\normalsize

The multi-product monopolist outcome faces unique changes as a result of heterogeneous attribute costs. I will prove the following:\\

\textit{\textbf{Proposition 8}: Under good-level heterogeneity in attribute costs, a multi-product product-designing monopolist optimising its products given multiple attributes will sell only a single good; that which, given the attribute costs it faces, yields the greatest expected profit. No horizontal or vertical differentiation takes place.}\\

\textit{Proof}: Let attribute costs differ by good, rather than firm as before. Let us set the per-good attribute costs in terms of $\boldsymbol{s}_k$ first. As before:

\begin{equation}
    \frac{1}{2}\sum_{n=1}^N\boldsymbol{s}_n'C_n\boldsymbol{s}_n=\frac{1}{2}\sum_{k=1}^K\sum_{l=1}^K\boldsymbol{s}_k'D_{kl}\boldsymbol{s}_l
\end{equation}

with $D_{kl},\forall k,l=1,\ldots K$ a diagonal matrix which places the $C_{n,kl}$ element in the $n$-th diagonal. Then, from the expected profit function:

\begin{equation}
    \frac{\partial \mathbb{E}\Pi}{\partial \boldsymbol{s}_k}=-\frac{1}{2\phi}(b_k-\gamma_k\boldsymbol{s}_k'M^{-1}\boldsymbol{\delta})M^{-1}\boldsymbol{\delta}-D_{kk}\boldsymbol{s}_k-\sum_{l\neq k}^KD_{kl}\boldsymbol{s}_l=0
\end{equation}

For $\lambda_k=-\frac{1}{2\phi}(b_k-\gamma_k\boldsymbol{s}_k'M^{-1}\boldsymbol{\delta}),\ \forall k$, the following optimal orientation rule is then true:

\begin{equation}\label{equivv}
    \sum_{l=1}^KD_{kl}\boldsymbol{s}_l=\lambda_k\boldsymbol{y}\qquad\Leftrightarrow\qquad C_n\boldsymbol{s}_n=y_n\boldsymbol{\lambda},\qquad \forall k=1,\ldots,K,\quad n=1\ldots,N
\end{equation}

with $\boldsymbol{\lambda}=(\lambda_1\ \ldots\ \lambda_K)'$ and $\boldsymbol{y}=M^{-1}\boldsymbol{b}$. Row by row, the following equivalence arises:

\begin{equation}
    C_n\boldsymbol{s}_n=y_n\boldsymbol{\lambda}\qquad\Rightarrow\qquad C_1\boldsymbol{s}_1\ ||\ \ldots\ ||\ C_N\boldsymbol{s}_N,\quad \forall n=1,\ldots,N
\end{equation}

This at first may appear to suggest that multi-product monopolists may deliver horizontal differentiation. Yet additional calculations reveal a different outcome. Note first that the LHS of the first equivalence can be re-written as follows:

\begin{equation}
    \sum_{l=1}^KD_{kl}\boldsymbol{s}_l=\lambda_k\boldsymbol{y}\qquad\Leftrightarrow\qquad\boldsymbol{s}_k=A_k(\boldsymbol{\lambda})\boldsymbol{y}
\end{equation}

for $A_k(\boldsymbol{\lambda})$ a diagonal matrix with diagonal elements $\{C_1^{-1}\boldsymbol{\lambda}\}_k,\ \ldots,\ \{C_N^{-1}\boldsymbol{\lambda}\}_k$ , $\forall k=1,\ldots,K$. Then, the following must be true:

\begin{equation}
    \begin{aligned}
        &\boldsymbol{\delta}=S\boldsymbol{b}\\
        \Leftrightarrow\ &\boldsymbol{\delta}=\sum_{k=1}^Kb_k\boldsymbol{s}_k\\
        \Leftrightarrow\ &\boldsymbol{\delta}=\sum_{k=1}^Kb_kA_k(\boldsymbol{\lambda})\boldsymbol{y}\\
        \Leftrightarrow\ &\Sigma\boldsymbol{y}=\sum_{k=1}^Kb_kA_k(\boldsymbol{\lambda})\boldsymbol{y}\\
        \Leftrightarrow\ &(I+\sum_{k=1}^K\gamma_k\boldsymbol{s}_k\boldsymbol{s}_k')\boldsymbol{y}=\sum_{k=1}^Kb_kA_k(\boldsymbol{\lambda})\boldsymbol{y}\\
        \Leftrightarrow\ &\boldsymbol{y}+\sum_{k=1}^K\gamma_k\boldsymbol{s}_k(\boldsymbol{s}_k'\boldsymbol{y})=\sum_{k=1}^Kb_kA_k(\boldsymbol{\lambda})\boldsymbol{y}\\
        \Leftrightarrow\ &\boldsymbol{y}+\sum_{k=1}^K\gamma_k(\boldsymbol{y}'A_k(\boldsymbol{\lambda})\boldsymbol{y})A_k(\boldsymbol{\lambda})\boldsymbol{y}=\sum_{k=1}^Kb_kA_k(\boldsymbol{\lambda})\boldsymbol{y}\\
    \end{aligned}
\end{equation}

Keep in mind that, from the earlier definition for $\lambda_k,\ \forall k=1,\ldots, K$:$b_k-\gamma_k\boldsymbol{y}'A_k(\boldsymbol{\lambda})\boldsymbol{y} = b_k-\gamma_k\boldsymbol{s}_k'\boldsymbol{y} =-2\phi\lambda_k$. Then:

\begin{equation}
\begin{aligned}
    &1+\sum_{k=1}^K\gamma_k{\{C_n^{-1}\boldsymbol{\lambda}\}_k}(\boldsymbol{y}'A_k(\boldsymbol{\lambda})\boldsymbol{y})=\sum_{k=1}^Kb_k{\{C_n^{-1}\boldsymbol{\lambda}\}_k}\\
    \Leftrightarrow&\sum_{k=1}^K(b_k-\gamma_k(\boldsymbol{y}'A_k(\boldsymbol{\lambda})\boldsymbol{y})){\{C_n^{-1}\boldsymbol{\lambda}\}_k}=1\\
    \Leftrightarrow&\sum_{k=1}^K\lambda_k{\{C_n^{-1}\boldsymbol{\lambda}\}_k}=\boldsymbol{\lambda}'C_n^{-1}\boldsymbol{\lambda}=-\frac{1}{2\phi}
\end{aligned}
\end{equation}

Notably, this condition will only be satisfied for two goods simultaneously if:

\begin{equation}
\begin{aligned}
    &\boldsymbol{\lambda}'(C^{-1}_n-C^{-1}_m)\boldsymbol{\lambda}=0\\
\end{aligned}
\end{equation}

This is ultimately a razor-thin margin. Hence, in general, only one good will satisfy it, meaning that the monopolist will only sell one good. Which good that will be will depend on which, if chosen, maximises expected profits. Heterogeneous costs have effectively introduced an incentive that breaks the indifference as to the distribution of attributes across goods sold by the monopolist. No horizontal differentiation takes place. \qedwhite\\

Lastly, note how, under firm- and product-level heterogeneous costs, as $M\rightarrow I$, $\boldsymbol{\lambda}=\boldsymbol{b}-\Gamma S'M\boldsymbol{\delta}\approx\boldsymbol{b}$, matching the approximate orientation rule which arises from the single-product firms case under the same conditions.\\

% In Section 3.4.3, we discussed what optimal attribute angles under symmetric competition look like. Turns out, this section provides a natural implication for that setting. If $C_n=C,\ \forall n$, because $C$ is invertible, it must be the case that $\boldsymbol{s}_1||\ldots||\boldsymbol{s}_N$. This also lets us rule out anti-parallel attributes entirely; as argued above, we cannot have two $\boldsymbol{s}_i$ and $\boldsymbol{s}_j$, for goods $i$ and $j$ respectively, be anti-parallel while simultaneously keeping positive $\delta$. Having ruled out anti-parallel attributes for any $C$ (not just small $C$), this is sufficient to confirm that $\boldsymbol{s}_i'\boldsymbol{s}_j=1,\ \forall i,j=1,\ldots N$, and therefore no horizontal competition takes place in equilibrium for the monopolist setting.

\section{Product Entry and Market Outcomes}

To complete this paper, I tackle the matter of market outcome prediction for hypothetical new goods. This section does not provide an exhaustive exploration of product entry, but instead aims to respond directly to the point raised in Gandhi and Nevo (2021) regarding the ability of product-based models such as linear demand to predict demand from new goods. We will prove the following:\\

\textit{\textbf{Proposition 10}: Greater product similarity between an incumbent firm and an entrant lowers incumbent firm demand and profits; makes a potential entrant less likely to enter; and has a minimal effect on consumer surplus. Upon entry, the incumbent's optimal product design tilts away from the monopoly orientation rule and loads more heavily on the exact opposite strategic direction to that determined by the entrant's design.}\\

\textit{Proof}: Consider an incumbent and a new entrant, both single-product firms. Mathematically, we can show that demand for the incumbent's good is as follows:

\begin{equation}
    q_{Inc}=\frac{m_{Ent}[(2m_{Inc}m_{Ent}-c^2)\delta_{Inc}-cm_1\delta_{Ent}]}{(m_{Inc}m_{Ent}-c^2)(4m_{Inc}m_{Ent}-c^2)}
\end{equation}

% \begin{equation}
%     q_{Ent}=\frac{m_{Inc}[(2m_{Inc}m_{Ent}-c^2)\delta_{Ent}-cm_{Ent}\delta_{Inc}]}{(m_{Inc}m_{Ent}-c^2)(4m_{Inc}m_{Ent}-c^2)}
% \end{equation}

for $m_i=1+\boldsymbol{s}_i'\Gamma \boldsymbol{s}_i$ and $c=\boldsymbol{s}_{Inc}'\Gamma \boldsymbol{s}_{Ent}$, where $M=[m_1\ c;\ c\ m_2]$. Expected firm profit on the other hand are as follows:

\begin{equation}
    \mathbb{E}\pi_{Inc}=-\frac{m_{Ent}[(2m_{Inc}m_{Ent}-c^2)\delta_{Inc}-cm_1\delta_{Ent}]^2}{\phi(m_{Inc}m_{Ent}-c^2)(4m_{Inc}m_{Ent}-c^2)^2}
\end{equation}

% \begin{equation}
%     \mathbb{E}\pi_{Ent}=-\frac{m_{Inc}[(2m_{Inc}m_{Ent}-c^2)\delta_{Ent}-cm_{Ent}\delta_{Inc}]^2}{\phi(m_{Inc}m_{Ent}-c^2)(4m_{Inc}m_{Ent}-c^2)^2}
% \end{equation}

At $c=0$, both firms behave as monopolists, and entry makes no difference. Consider instead the case where $c$ is perturbed such that it reflects greater similarity in product characteristics. Then, incumbent 

\begin{equation}
    \frac{\partial q_{Inc}}{\partial c}\approx -\frac{\delta_{Ent}}{4m_{Inc}m_{Ent}}
\end{equation}

\begin{equation}
    \frac{\partial \mathbb{E}\pi_{Inc}}{\partial c}\approx \frac{\delta_{Inc}\delta_{Ent}}{\phi4m_{Inc}m_{Ent}}
\end{equation}

This suggests that around $c=0$, greater characteristic overlap lowers both the incumbent firm’ quantities and profits linearly. Zero-overlap entry is innocuous to the incumbent, while entry by a similar firm has a negative first-order effect that is linear in similarity. It can be shown that similarity has the identical effect for the entrant, meaning that similarity with incumbent goods makes entry less likely.\\

What about consumer welfare? It can be shown that similarity plays a minimal role, with effects driven primarily by initial product utility (much like Logit, consumer surplus increases by construction with product entry).

\begin{equation}
    CS=\frac{1}{2}\boldsymbol{q}^{*'}M^{-1}\boldsymbol{q}^{*} \qquad\Rightarrow\qquad \frac{\partial}{\partial c}CS(c)\bigg|_{c=0}\approx0
\end{equation}

This completes our proof.\qedwhite\\

The model we have discussed in this paper also allows for further discussion on how new products, upon entry, affect optimal product design. From (\ref{parallelsingprod}), pertaining to optimal attribute orientation for single-product firms under attribute cost heterogeneity, we know that optimal product design for good $n$, $\boldsymbol{s}_n^*$ is that such that $\boldsymbol{s}_n^*$ is a weighted combination of $N$ directions: $\boldsymbol{b}$, as well as every other $\boldsymbol{s}_i$ vector. Furthermore, from the above, we know that $\frac{\partial \mathbb{E}\pi_{Inc}}{\partial M_{Inc,Ent}}=\frac{\partial \mathbb{E}\pi_{Inc}}{\partial c}<0$. We would therefore expect entry to tilt design away from the monopoly direction $C^{-1}\boldsymbol{b}$ and toward a strategic direction determined by rival design. The negative sign implies that, for greater similarity between the incumbent and the new entrant's goods, the incumbent loads more heavily on desiging \textit{away} from the entrant and less heavily towards $C^{-1}\boldsymbol{b}$. In other words, a new good will likely make its rivals simultaneously more opposed to itself and less like weighted attribute utility.\\

Much has been said in the product design literature on the concept of \textit{niche} designs (see Bar-Isaac, Caruana, and Cunat, 2021, for a recent assessment). In particular, recent work has suggested that "extreme" designs may be an early strategy upon entry for a firm, with a greater focus on a broader audience over time reflecting something closer to equilibrium behaviour (Gong, 2021). This paper therefore provides one explanation - a niche design may make a firm more likely to enter a market, even if the equilibrium design is eventually of broader interest. While outside the scope of this paper, another way niche products these may arise in the model is if the assumption of a single representative consumer is broken, i.e. when consumers disagree as to what goods are substitutes and complements and to what extent. Further research is needed to determine if and when such degree of consumer heterogeneity takes place, since it does not necessarily follow from the existing literature on consumer segmentation/clustering.\\

Also outside the scope of this paper is a broader discussion on innovation. Innovation is often split between a new product (e.g. a new characteristic) or a new process (impacting production costs). It may be asserted that the model in this paper can help highlight another source of product innovation - a new technology mix. A new product, e.g. a phone with the same size but lighter, or an apple that stays ripe for longer without losing taste, changes the perceived relationship between product characteristics. For example, a new good which allows the same volume at much lower mass than previously observed changes the density distribution across the product category, impacting the design equilibrium. Further research should help uncover unique implications of this.\\

\section{Numerical Example}

To complete this paper, we briefly consider a working numerical example of the model described above. Consider a setting with two single-product price-setting firms, one selling a premium phone, and another a "budget" phone. These phones have two observable characteristics: battery life and camera quality. We can write the respective product characteristics matrix $X$ as follows:

\begin{equation}
    X=\begin{pmatrix}20&16\\10&7\end{pmatrix}=\bigg(\frac{1}{\sqrt{5}}\begin{pmatrix}2&-1\\1&2\end{pmatrix}\bigg)\bigg(\sqrt{5}\begin{pmatrix}10&7.8\\0&-0.4\end{pmatrix}\bigg)=ZR
\end{equation}

The premium phone has almost twice the "points" on both characteristics. "Points" here may refer to any unit we may wish to use for these characteristics: e.g. battery capacity in mAh; or camera resolution in megapixels. Let the former be the most relevant to a representative consumer:

\begin{equation}
\boldsymbol{\beta}
=\begin{pmatrix}0.10\\0.05\end{pmatrix}
\quad\Rightarrow\quad
\mathbb{E}\boldsymbol{\delta}=X\boldsymbol{\beta}=\begin{pmatrix}2.8\\1.35\end{pmatrix}
\end{equation}

We will also define the Hessian matrix of the representative consumer's utility function, $M$:

\begin{equation}
M=\begin{pmatrix}3.5&1.5\\1.5&3.5\end{pmatrix}
= (ZU)\Gamma (ZU)'+I=
S\Gamma S' + I
\end{equation}
\begin{equation}
S=\frac{1}{\sqrt2}\begin{pmatrix}1&1\\1&-1\end{pmatrix},\ 
\Gamma=\begin{pmatrix}4&0\\0&1\end{pmatrix}
,\ U=\frac{1}{\sqrt{10}}\begin{pmatrix}3&1\\-1&3\end{pmatrix}\end{equation}

Despite the very different levels of their characteristics, to the point of being vertically differentiated in our observed characteristics, these goods appear to be imperfect substitutes with similar demand slopes. In re-writing $M$ in terms of $\Gamma$ and $S$, the model concilliates this paradox by weighing heavily the competitive salience of the first attribute in $S$, where both goods seem to load similarly, while weighing the competitive salience of the second attribute, where they differ more substantially, more lightly.\\

% Interpreting the attributes is far harder than interpreting characteristics. The Gram-Schmidt process, a common approach to QR decomposition, defines the $n$-th attribute in a set as approximately the $n$-th characteristic net of their correlation to the $n-1$ previous characteristics. In this sense, one may think of the two attributes above as (i) battery life and (ii) camera net of battery-life-related qualities. For example, low power can limit features like flash, slow down computational photography processing, or cause earlier camera shutdowns. In that sense, our budget phone can be interpreted as relying relatively more on its battery life over independently investing into camera quality. This is nonetheless unimportant, as optimal product designing works regardless of how characteristics relate to any of many orthonormal attribute characterisations.\\

From the above, we can produce the implied attribute utility $\boldsymbol{b}$:

\begin{equation}
\boldsymbol{b}=S^{-1}\mathbb{E}\boldsymbol{\delta}= \begin{pmatrix}2.93\\1.03\end{pmatrix}
\end{equation}

To account for the significant differences in the number of points across the two goods, we pre-calculate their implied attribute costs as an "attribute budget". In practice, this means our results will resemble the setting with attribute cost heterogeneity. Let $\phi=-1$. The results are as follows:

\begin{equation}
    X^*_{mono,excl}=
    \begin{bmatrix}
        27.58 &   21.15\\
        11.47 &  10.32\\
    \end{bmatrix}
    \qquad 
    X^*_{mono,non-excl}=
    \begin{bmatrix}
        33.04  & 26.73\\
        0  & 0\\
    \end{bmatrix}
\end{equation}
\begin{equation}
    X^*_{sing,excl}=
    \begin{bmatrix}
        27.58 &  21.15\\
        11.47 &  10.32
    \end{bmatrix}
    \qquad 
    X^*_{sing,non-excl}=
    \begin{bmatrix}
        30.98 &  25.07\\
        11.39 &   9.20
    \end{bmatrix}
\end{equation}

The differences in cost across goods lead the monopoly towards the one-good outcome we discussed in a previous section; firm-level heterogeneity as a multiplicative of a common $C$ on the other hand only affected attribute intensity - attributes are parallel as predicted by the cost homogeneity setup. The difference lies in that the monopoly is very sensitive to any heterogeneity, whereas individual firms require greater diversity of cost profiles to deliver greater variety of orientations.\\

Average expected profit per good are as follows:
\begin{equation} 
\begin{aligned}
    \mathbb{E}\overline\pi^*_{mono,non-excl}>\mathbb{E}\overline\pi^*_{mono,excl}=\mathbb{E}\overline\pi^*_{sing,excl}>\mathbb{E}\overline\pi^*_{sing,non-excl}
\end{aligned}    
\end{equation}
\begin{equation} \nonumber
\begin{aligned}
    \Leftrightarrow\ 0.3035>0.2733=0.2733>0.2212
\end{aligned}    
\end{equation}

As discussed in a previous section, outcomes are identical in the two attribute-exclusivity settings. The unique difference is that, for a monopolist, this is a constraint; for single-product firms, it constitutes an opportunity for collusion.\\

\section{Discussion and Conclusion}

This paper has shown how a fundamentally product-based demand system as linear demand is can be extended into a fully characteristics-based framework without losing the tractability that makes it analytically useful. By allowing product characteristics to shape both consumers’ valuations and the competitive relationships across goods, the paper demonstrates - in a precursor to future work - that linear demand can speak directly to questions on endogenous product design and entry. More broadly, the paper establishes that the model remains workable in settings with any finite number of goods, firms, and characteristics, and under both symmetry and asymmetry in cost structures.\\

I have not analysed the matter of consumer heterogeneity in this paper. It could be presumed this is due to some limitation of the model - this is not so. Once more than one consumer is considered, we move outside the realm of the simple proofs and closed-form solutions. Yet it has been convincingly demonstrated here that numerical solutions can converge fast in a linear demand model. Presumably, in a more complete model as in Rodrigues (2026), firms could study corner solutions, adjusting prices and characteristics to segment consumers. What this may look like mathematically would require extending this paper beyond what was its original scope: to demonstrate that linear demand is as robust an approach to study optimal product design and demand for new products as any existing alternative method.

\newpage
\section{References}
\small\singlespacing

Akerlof, G. (1970). The Market for "Lemons": Quality Uncertainty and the Market Mechanism. The Quarterly Journal of Economics, 84(3), 488–500.\\

Amir, R., Erickson, P., and Jin, J. (2017). On the microeconomic foundations of linear demand for differentiated products. Journal of Economic Theory, 169, 641-665.\\

Bar-Isaac, H., Caruana, G., and Cuñat, V. (2012). Search, Design, and Market Structure. American Economic Review, 102(2), 1140–60.\\

Bar-Isaac, H., Caruana, G., and Cuñat, V. (2023). Targeted Product Design. American Economic Journal: Microeconomics, 15(2), 157–86.\\

Bordalo, P., Gennaioli, N., and Shleifer, A. (2013). Salience and Consumer Choice. Journal of Political Economy, 121(5), 803-843.\\

Butters, G. (1977). Equilibrium distributions of sales and advertising prices. The Review of Economic Studies, 44(3), 465–491.\\

Campbell, J., and Viceira, L. (2002). Strategic Asset Allocation: Portfolio Choice for Long-Term Investors. Oxford: Oxford University Press. \\

Chen, Y., and Riordan, M. (2007). Price and variety in the spokes model. The Economic Journal, 117(522), 897–921.\\

Choné, P., and Linnemer, L. (2020). Linear demand systems for differentiated goods: Overview and user’s guide. International Journal of Industrial Organization, 73, 102663.\\

Darton, R. (1980). Rotation in factor analysis. Journal of the Royal Statistical Society Series D: The Statistician, 29(3), 167–194.\\

d’Aspremont, C., Gabszewicz, J., and Thisse, J. (1979). On Hotelling’s “stability in competition”. Econometrica, 47(5), 1145–1150.\\

Economides, N. (1993). Quality variations in the circular model of variety-differentiated products. Regional Science and Urban Economics, 23(2), 235-257.\\

Fama, E., and French, K. (1993). Common risk factors in the returns on stocks and bonds. Journal of Financial Economics, 33(1), 3-56. \\

Golub, G., and van Loan, C. (1983). Matrix computations. Johns Hopkins University Press: Baltimore.\\

Gong, Z. (2021). Growing Influence. Working Paper.\\

Gorman, W. (1959). Separable Utility and Aggregation. Econometrica, 27(3), 469-481.\\

Hoberg, G., and Phillips, G. (2016). Text-Based Network Industries and Endogenous Product Differentiation. Journal of Political Economy, 124(5), 1423-1465.\\

Hotelling, H. (1929). Stability in competition. The Economic Journal, 39(153), 41-57.\\

Johnson, J., and Myatt, D. (2006). On the Simple Economics of Advertising, Marketing, and Product Design. American Economic Review, 96(3), 756–784.\\

Koijen, R. and Motohiro, Y. (2019). A Demand System Approach to Asset Pricing. Journal of Political Economy, 127(4), 1475-1515.\\

Kreps, D., and Scheinkman, J. (1983). Quantity Precommitment and Bertrand Competition Yield Cournot Outcomes. Bell Journal of Economics, 14(2), 326-337.\\

Lancaster, K. (1966). A new approach to consumer theory. Journal of Political Economy, 74, 132-157.\\

Lancaster, K. (1990). The economics of product variety: A survey. Marketing Science, 9(3), 189–206. \\

Levitan, R., and Shubik, M. (1972). Price duopoly and capacity constraints. International Economic Review, 13(1), 111–122.\\

Miyashita, M. (2026). Characteristics Design:A Hedonic Approach to Optimal Product Differentiation. Working paper.\\

Moorthy, K. (1988). Product and Price Competition in a Duopoly. Marketing Science, 7(2), 141-168.\\

Motta, M. (1993). Endogenous Quality Choice: Price vs. Quantity Competition. The Journal of Industrial Economics, 41(2), 113-131. \\

Mussa, M., and Rosen, S. (1978). Monopoly and product quality. Journal of Economic Theory, 18(2), 301-317. \\

Peitz, M. and Sato, S. (2025). Asymmetric Platform Oligopoly. The RAND Journal of Economics, 57(1), 60-77. \\

Pellegrino, B. (2025). Product differentiation and oligopoly: A network approach. American Economic Review, 115(4), 1170-1225.\\

Perloff, J., and Salop, S. (1985). Equilibrium with product differentiation. The Review of Economic Studies, 52(1), 107–120.\\

Rhodes, A., and Zhou, J. (2024). Personalized Pricing and Competition. American Economic Review, 114(7), 2141–70.\\

Rodrigues, A. (2026). Consumer choice over shopping baskets: a linear demand approach. arXiv preprint arXiv:2511.11846.\\

Salop, S. (1979). Monopolistic competition with outside goods. The Bell Journal of Economics, 10(1), 141-156.\\

Shaked, A., and Sutton, J. (1982). Relaxing price competition through product differentiation. The Review of Economic Studies, 49(1), 3–13.\\

Sherman, J., and Morrison, W. (1949). Adjustment of an Inverse Matrix Corresponding to Changes in the Elements of a Given Column or a Given Row of the Original Matrix. Annals of Mathematical Statistics. 20, 621.\\

Spence, A. (1975). Monopoly, quality, and regulation. The Bell Journal of Economics, 6(2), 417–429.\\

Tirole, J. (1988). The Theory of Industrial Organization. MIT Press: Cambridge.\\

Ushchev, P., and Zenou, Y. (2018). Price competition in product variety networks. Games and Economic Behavior, 110, 226-247.\\

Vives, X. (1987). Small Income Effects: A Marshallian Theory of Consumer Surplus and Downward Sloping Demand. The Review of Economic Studies, 54(1), 87-103.\\

von Ungern-Sternberg, T. (1988). Monopolistic Competition and General Purpose Products. The Review of Economic Studies, 55(2), 231-246.\\

Voelkening, D. (2026). Product Design in an Hedonically Differentiated Duopoly. Working paper. \\

Woodbury, M. (1950). Inverting modified matrices (Memorandum Rept. 42, Statistical Research Group). Princeton: Princeton University.\\

Zeitlin, J., and Rangoni, B. (2025). How the European Union reconciles uniform regulation with legitimate diversity: towards a tighter experimentalist governance architecture. Journal of European Public Policy, 1–29.\\

\normalsize %\doublespacing

\newpage

\appendix

\section{Appendix}
\small\singlespacing

\subsection{Proofs}

\textit{Lemma 1.} Note that $\mathrm{span}(X)=\mathrm{span}(Z)$. Let $Q = [Z\ Z_\perp]$, an orthonormal basis composed of $Z$ and its orthogonal complement $Z_\perp$. Orthogonal complements span the vector space outside the range of a given matrix; i.e. $Z_\perp$ spans the set of all vectors that are orthogonal to every vector in $Z$. For $M$ a full-rank matrix (necessary and sufficient for a well-behaved utility function), our assertion that it is partially a function of $Z$ implies that the remaining variation is represented by coordinates in $Z_\perp$.\\

Let $\hat M=E(M|X)$ and:

\begin{equation}
    \tilde M = Q\hat M Q'
\end{equation}

The $\tilde M$ matrix is $\hat M$ but expressed in the $Q$-basis. Being symmetric, any $\tilde M$ can be written as a block matrix with the following structure:\\

\begin{equation}
    \tilde M = Q\hat M Q'= \begin{bmatrix}
        D & B\\ B' & C
    \end{bmatrix}
\end{equation}

with $D\in\mathbb{R}^{K\times K}$, $B=B'\in \mathbb{R}^{K\times N-K}$, and $C\in\mathbb{R}^{(N-K)\times (N-K)}$. This follows from the fact that $\tilde M$ maintains the symmetry properties of $\hat M$.\\

Now, without any information outside the span of $Z$, i.e. in $\mathrm{span}(Z_\perp)$, we may wish to remain agnostic about what is an appropriate approximation for that portion of $M$. To formalise that ignorance, we may require that $\tilde M$ be invariant to orthogonal transformations acting only on the complement subspace. In other words, we adjust $\tilde M$ so that it is isotropic outside the span of $Z$. Isotropy can be defined as follows:

\begin{equation}
    \begin{bmatrix} I_K & 0\\ 0 & H \end{bmatrix}\hat M
    \begin{bmatrix} I_K & 0\\ 0 & H
    \end{bmatrix}
    =
    \tilde M
\end{equation}

for any random matrix $H\in\mathbb{R}^{(N-K)\times (N-K)}$. The idea here is that any misspecification in the unobserved component $\mathrm{span}(Z_\perp)$ will be less likely to dominate $\hat M$ if all possibilities within that component are treated in the same way. Matrix $H$ is used here to show the kind of transformation we want $M$ to be unaffected by - the more affected the expression is by $R$, the more likely it is that misspecification of $\hat M$ in the unobserved component span dominates the matrix, obfuscating the part of $M$ that we do know and are trying to model (i.e. the part dependent on characteristics). From before, we have:

\begin{equation}
    \begin{bmatrix} I_K & 0\\ 0 & H \end{bmatrix}
    \begin{bmatrix}
        D & B\\ B' & C
    \end{bmatrix}
    \begin{bmatrix} I_K & 0\\ 0 & H
    \end{bmatrix}
    =
    \begin{bmatrix}
        D & BH'\\ HB' & HCH'
    \end{bmatrix}
\end{equation}

we may only have isotropy outside $\mathrm{span}(Z)$ if (i) $BH'=B$ and (ii) $HCH'=C$. Take e.g. $H=-I$. Then $BH'=B(-I)=-B$. In other words, a non-zero matrix $B$ is a contradiction: $B=0$. It is similarly straightforward to show that $C=\rho I$ by first demonstrating that it must be diagonal (or else a reflection matrix can flip the sign of any off-diagonal coordinate and break the equality) and all its diagonal elements must be identical (or else a permutation matrix can change the order of the rows/columns and break the equality).\\

With $B=0$ and $C=\rho I$, restate $\tilde M = Q\hat M Q'$ in terms of $\hat M$:

\begin{equation}
        \hat M = Q'\tilde M Q= Q'\begin{bmatrix}
        D & 0\\ 0 & \rho I
    \end{bmatrix}Q
    =
    ZAZ'+\rho(I-ZZ')
\end{equation}

This completes the proof.\qedwhite\\

\textit{Lemma 2.} There is always a $\boldsymbol{b}$ such that $Z\boldsymbol{\tilde\beta}=ZU\boldsymbol{b}\Leftrightarrow Z(\boldsymbol{\tilde\beta}-U\boldsymbol{b})=0$ if $U$ is invertible; then $\boldsymbol{b}=U^{-1}\boldsymbol{\tilde\beta}$. This will be the case if $A$, from which $U$ is derived, is symmetric, which it is by construction as $M$ is positive definite. Then $U$, as an orthogonal matrix of $A$'s eigenvectors, is invertible, and we may re-write $\boldsymbol{\delta}=S\boldsymbol{b}+\boldsymbol{v}$, for $S=ZU$. If $\boldsymbol{\beta}>0$, note that we can always define a $\boldsymbol{b}>0$, as eigenvectors are defined up to a sign - we can change the signs of the columns of $U$ as needed to achieve this.\qedwhite 
%Note that if $\boldsymbol{\beta}>0$, we may always assume $\boldsymbol{b}>0$. This is because $R>0$ and because we may flip the sign of any given column of $U$ without loss of generality, as eigenvectors are determined only up to sign.
\\

\textit{Lemma 3.} In general, when writing $\mathbb{E}\Pi$, all terms related to $\boldsymbol{v}$ can be ignored. We can see this from $\mathbb{E}\Pi=(-1/(4\phi))\boldsymbol{\delta}'M^{-1}\boldsymbol{\delta}=(-1/(4\phi))(X\boldsymbol{\beta})'M^{-1}(X\boldsymbol{\beta})+2(-1/(4\phi))(X\boldsymbol{\beta})'M^{-1}\boldsymbol{v}+(-1/(4\phi))\boldsymbol{v}'M^{-1}\boldsymbol{v}$. The term in the middle drops out as  $\mathbb{E}\boldsymbol{v}=0$. The second term is constant in S: $\mathbb{E}(\boldsymbol{v}'M^{-1}\boldsymbol{v}|X)=\mathbb{E}(\mathrm{tr}(\boldsymbol{v}'M^{-1}\boldsymbol{v}))|X)=\mathbb{E}(\mathrm{tr}(M^{-1}\boldsymbol{v}'\boldsymbol{v})|X)=\mathrm{tr}(M^{-1}\mathbb{E}(\boldsymbol{v}\boldsymbol{v}')|X)$; our unit variance assumption (we may allow this up to a constant, only homoskedasticity is required) together with $\mathbb{E}\boldsymbol{v}=0$ implies $\mathrm{tr}(M^{-1}\mathbb{E}(\boldsymbol{v}\boldsymbol{v}')|X)=tr(M^{-1})=\sum_{k=1}^K1/\gamma_k$, which is independent from $S$.\qedwhite

\subsection{Finding $S$ when $X$ is not observed but $M$ is}

Just as we can produce a positive definite matrix approximately from assuming it is partially a function of $X$ using $X$'s orthogonal representation, we can take advantage of the spectral decomposition of a known measure of competition to produce the relevant attributes which drive said competition. Note that, for any matrix, we can obtain the following spectral decomposition:

\begin{equation}\nonumber
    M=\sum_{i=1}^N\lambda_i\boldsymbol{u}_i\boldsymbol{u}_i',\ \lambda_1\geq\lambda_2\geq\ldots\geq\lambda_N\geq0
\end{equation}

For an $N\times N$ matrix $\Sigma$, we can write:

\begin{equation}\nonumber
    \begin{aligned}
    M& = \sum_{n=1}^N \lambda_n \boldsymbol{u}_n \boldsymbol{u}_i'\\
    &=\sum_{n=1}^N  (\lambda_n-\rho) \boldsymbol{u}_n \boldsymbol{u}_n'+\sum_{i=1}^N \rho\cdot \boldsymbol{u}_i \boldsymbol{u}_i'\\
    & =\sum_{n=1}^N  \gamma_n\boldsymbol{u}_n \boldsymbol{u}_n+\rho I_N
    \end{aligned}
\end{equation}

for $\gamma_n=\lambda_n-\rho,\ \forall n=1,\ldots,N$. The $\boldsymbol{u}$ here effectively operate as $\boldsymbol{s}$ in the earlier specification. To keep the assumption of $\gamma_n>0,\ \forall n$, however, the above equation will only be approximately true; if $\exists \lambda_k:\lambda_k<\rho$, then we must effectively set $\lambda_k=\rho$, as otherwise $M$ will not be positive definite as required. This will nonetheless be at least as good as in the case where we have $X$ but not $M$, where some of the attributes for which $\gamma>0$ may not be observed.\\

\subsection{Connection between $M$ and $\mathbb{E}(\boldsymbol{\delta}\boldsymbol{\delta}')$}

Our $\hat M$ measure can be shown to be proportional to the second (raw) moment of $\boldsymbol{\delta}$, $E(\boldsymbol{\delta}\boldsymbol{\delta}')$. Under the assumptions we have placed earlier on each term in $\boldsymbol{\delta}$:

\begin{equation}\nonumber
    \mathbb{E}(\boldsymbol{\delta}\boldsymbol{\delta'}) = \mathbb{E}(S\boldsymbol{b}\boldsymbol{b}'S')+\mathbb{E}(\boldsymbol{vv'})=\sum_{k=1}^{K}\gamma_k\boldsymbol{s}_k\boldsymbol{s}_k'+I_N
\end{equation}

for $\gamma_k= b_k^2$. In this sense, $\gamma_k,\ \forall k$ can be framed within a representative consumer setup as a measure of dispersion in mean utility from attribute $k$ along the consumer population. In general, the paper assumes $\gamma_k\neq b_k^2$.\\

It is well-understood in the Statistics literature on common factor models - from which this paper draws heavily - that any vector which can be described as a linear function of factors (such as $\boldsymbol{\delta}$ and $S$ above) can, under certain properties, allow the variance-covariance matrix of said factors be described as above (see e.g. Darton, 1980). It is also well-known in the Finance literature that quadratic utility is consistent with mean–variance analysis (see e.g. Ross, 1977; Fama and French, 1993), with the Hessian matrix $M$ encoding expected asset return variance. The above results from the interaction between these two fields, which I repurpose here for use in empirical microeconomics (and more specifically industrial organisation).

\subsection{Exploratory analysis of empirical testability}

In Rodrigues (2026), an approach is introduced for estimating linear demand. In this paper, we have used the same model with some adjustments. In this Appendix, I explore how such adjustments impact the estimation of aggregate linear demand models. More specifically, I make a first foray into the matter of estimating the parameters exclusive to this paper, verifying whether a minimal condition for estimation - the well-behavedness of the model, in respect of these parameters, within the context of linear regression - is satisfied.\\

\subsubsection{Estimating $U$}

Throughout the paper, we have worked with $S=ZU$. However, while $Z$ may in general be thought of as observable, through QR decomposition on the matrix of observed characteristics $X$, $U$ is not. In this section, I break down how we may parameterise $U$ such that it is estimable via econometric methods. I propose the following specification for $U(\boldsymbol{\theta})$:

\begin{equation}
    U(\boldsymbol{\theta}) = \prod_{0\leq i<j\leq K} G(i,j,\theta_{ij})
\end{equation}
\vspace{0.7pt}\\
for $G(i,j,\theta_{ij})$ a Givens rotation (Golub and van Loan, 1983, p.240):\\

\begin{equation}
    G(i,j,\theta_{ij})=\begin{bmatrix}
        1 & \cdots & 0 & \cdots & \cdots &\cdots & 0 & \cdots & 0\\
        \vdots & \cdots & \vdots & \cdots & \vdots &\cdots &\vdots & \cdots & \vdots\\
        0 & \cdots & \cos(\theta_{ij}) & 0 &\cdots & \cdots & \sin(\theta_{ij}) & \cdots & 0\\
        \vdots & \cdots & \vdots & \cdots & \vdots &\cdots&  \vdots & \cdots & \vdots\\
        0 & \cdots & -\sin(\theta_{ij}) & 0 & \cdots &  \cdots & \cos(\theta_{ij}) & \cdots & 0\\
        \vdots & \cdots & \vdots & \cdots & \vdots &\cdots& \vdots & \cdots & \vdots\\
        0 & \cdots & 0 & \cdots & \cdots & \cdots & 0 & \cdots & 1\\        
    \end{bmatrix}
\end{equation}
\vspace{0.7pt}\\
with non-$\{0,1\}$ elements reflecting where the rows $i$ and $j$ (self-)intersect. A Givens rotation turns two coordinates and leaves all the others unchanged. For example, for a 2-by-2 $U(\boldsymbol{\theta})$, if $\boldsymbol{y}=U(\boldsymbol{\theta})'\boldsymbol{x}$, the $\boldsymbol{y}$ is obtained by rotating $\boldsymbol{x}$ counter clockwise through an angle $\theta_{ij}$.\\

We can confirm $G(i,j,\theta_{ij})$ is orthogonal: $G' G=I$; $\det(G)=1$; and $G(i,j,\theta_{ij})^{-1}=G(i,j,-\theta_{ij})$. Furthermore, Givens rotations' flexibility allows us to, by chaining a small number of such two–coordinate rotations, get a rotation along multiple coordinates at once. Each factor uses a chosen coordinate pair $(i,j)$ and one angle $\theta_{ij}$.\\

If $Z=[\boldsymbol{z}_1,\dots,\boldsymbol{z}_K]$ has orthonormal columns (a basis for the span of $X$), we can turn those directions by $U(\boldsymbol{\theta})$:

\begin{equation}
    S = ZU(\boldsymbol{\theta}) = [\boldsymbol{s}_1,\dots,\boldsymbol{s}_K]
\end{equation}

and then proceed as we have so far. Because $U(\boldsymbol{\theta})$ is orthogonal, the $\boldsymbol{s}_k$ remain orthonormal.\\

Before moving on to studying the well-behaveness of a regression equation relative to the $\boldsymbol{\theta}$ parameters, note the following. Firstly, the order of the multiplication in $U(\boldsymbol{\theta})$ matters: $G(i,k,\theta_{ik})G(i,j,\theta_{ij})\neq G(i,j,\theta_{ij})G(i,k,\theta_{ik}),\ \forall i,j,k$, so they do not commute. For empirical estimation of our model, however, this does not matter. The $\boldsymbol{\theta}$ parameters are meant only to provide the necessary fit for how our product characteristics $X$ enter $M$. Secondly, the total number of parameters, for $K$ the number of product attributes, is up to $\frac{K(K-1)}{2}$. This depends only on the number of attributes, not of goods, and therefore likely scales well where $K<N$.\\

Let us now consider whether, for $U(\boldsymbol{\theta})$ defined via Givens rotations, our regression equation is well-behaved in the $\boldsymbol{\theta}$ parameters. Let:

\begin{equation}
M(\boldsymbol{\theta}) = I + \sum_{k=1}^K \gamma_k\boldsymbol{s}_k(\boldsymbol{\theta}) \boldsymbol{s}_k(\boldsymbol{\theta})'
\end{equation}

Then:

\begin{equation}
M^{-1}(\boldsymbol{\theta}) = I - \sum_{k=1}^K \alpha_k \boldsymbol{s}_k(\boldsymbol{\theta}) \boldsymbol{s}_k(\boldsymbol{\theta})'
\end{equation}

via orthonormality in the columns of S and the Sherman-Woodbury formula, with $\alpha_k=\frac{\gamma_k}{1+\gamma_k}\in[0,1)$ and $\gamma_p\neq\gamma_q,\ \forall p,q$. I will prove well-behavedness in $M^{-1}(\boldsymbol{\theta})$ without loss of generality, allowing researchers to choose whichever alternative works best for their empirical specification.\\

Consider the behaviour of the model with regards to parameter $\theta$, which changes the angle between the pair of attribute vectors $\boldsymbol{s}_p(\theta)$ and $\boldsymbol{s}_q(\theta))$. We can write how these vectors depend on $\theta$ as follows:

\begin{equation}
    [\boldsymbol{s}_p(\theta)\ \boldsymbol{s}_q(\theta)]=[\boldsymbol{s}_p\ \boldsymbol{s}_q]R(\theta)
    \qquad
    R(\theta)=\begin{bmatrix}\cos\theta&-\sin\theta\\ \sin\theta&\cos\theta\end{bmatrix}.
\end{equation}

The matrix $R(\theta)$ is known as a \textit{rotation matrix}. All other $\boldsymbol{s}_k$ are fixed. Let $\boldsymbol{y},\boldsymbol{x}\in\mathbb{R}^N$ be given, for $\boldsymbol{y}=M^{-1}\boldsymbol{x}$ a stylised version of the linear regression model of Rodrigues (2026). Decompose without loss of generality
$\boldsymbol{y}=\boldsymbol{y}_\perp+ y_p \boldsymbol{s}_p+y_q \boldsymbol{s}_q$, $\boldsymbol{x}=\boldsymbol{x}_\perp+ x_p \boldsymbol{s}_p+x_q \boldsymbol{s}_q$.\footnote{Any vector $\boldsymbol{w}$ can be decomposed into a portion within $\mathrm{span}(S)$ and another portion in said vector space's orthogonal complement: i.e. $\boldsymbol{w}=\sum_k w_k\boldsymbol{s}_k + \boldsymbol{w}_\perp$ with $w_k=\boldsymbol{s}_k'\boldsymbol{w}$ and $\boldsymbol{w}_\perp\perp\mathrm{span}\{\boldsymbol{s}_1,\dots,\boldsymbol{s}_K\}$, for any $\boldsymbol{w}\in R^N$.}
Define $\beta_k=1-\alpha_k=\frac{1}{1+\gamma_k}>0$,
$\beta_s=\frac{\beta_p+\beta_q}{2}$, and $\beta_d=\frac{\beta_p-\beta_q}{2}$.\\

Note then that:

\begin{equation}
\begin{aligned}
    &M(\theta)^{-1}\boldsymbol{x}_\perp=(I-\sum_{k=1}^K\alpha_k\boldsymbol{s}_k\boldsymbol{s}_k')\boldsymbol{x}_\perp=\boldsymbol{x}_\perp\\
    &M(\theta)^{-1}\boldsymbol{s}_j=(I-\sum_{k=1}^K\alpha_k\boldsymbol{s}_k\boldsymbol{s}_k')\boldsymbol{s}_j=\boldsymbol{s}_j-\alpha_j\boldsymbol{s}_j=\beta_j\boldsymbol{s}_j,\forall j\neq p,q
\end{aligned}
\end{equation}

This shows that (i) $M^{-1}(\theta)$ acts only upon the vectors in $\mathrm{span}(Z)$; and (ii) the regression model is unaffected by changes in $\theta$ for vectors that are not $p$ and $q$. In other words, we are able to isolate our focus on how the angle between $\boldsymbol{s}_p$ and $\boldsymbol{s}_q$ changes how these variables drive demand.\\

Let us then isolate the part of $M^{-1}$ which depends on these two variables and is in fact affected by their angle. A simplified $M^{-1}$ may be written as follows:

\begin{equation}
\begin{aligned}
M(\theta)^{-1}= &\ I-\alpha_p\boldsymbol{s}_p(\theta)\boldsymbol{s}_p(\theta)'-\alpha_q\boldsymbol{s}_q(\theta)\boldsymbol{s}_q(\theta)'\\
= &\ I-[\boldsymbol{s}_p(\theta)\ \boldsymbol{s}_q(\theta)]\begin{bmatrix}
    \alpha_p&0 \\ 0& \alpha_q \\
\end{bmatrix}[\boldsymbol{s}_p(\theta)\ \boldsymbol{s}_q(\theta)]'\\
= &\ I - [\boldsymbol{s}_p\ \boldsymbol{s}_q]R(\theta)\begin{bmatrix}
    \alpha_p&0 \\ 0& \alpha_q \\
\end{bmatrix}R(\theta)'[\boldsymbol{s}_p\ \boldsymbol{s}_q]'
\end{aligned}
\end{equation}

Writing $M^{-1}\boldsymbol{w}$, for $\boldsymbol{w}$ any vector in the span of vectors $\boldsymbol{s}_p$ and $\boldsymbol{s}_q$ (i.e. $\boldsymbol{w}=[\boldsymbol{s}_p\ \boldsymbol{s}_q]\boldsymbol{u}$, for $\boldsymbol{u}\in\mathbb{R}^2$) and applying the fact that the outcome variable $\boldsymbol{y}$ is only relevant to our purposes within the same vector span ($\boldsymbol{y}$ outside this span are unaffected by $\theta$):

\begin{equation}
\begin{aligned}
    M^{-1}(\theta)\boldsymbol{w}= &\ (I - [\boldsymbol{s}_p\ \boldsymbol{s}_q]R(\theta)\begin{bmatrix}
    \alpha_p&0 \\ 0& \alpha_q \\
\end{bmatrix}R(\theta)'[\boldsymbol{s}_p\ \boldsymbol{s}_q]')[\boldsymbol{s}_p\ \boldsymbol{s}_q]\boldsymbol{u}\\
    = &\ [\boldsymbol{s}_p\ \boldsymbol{s}_q](I_2-R(\theta)\begin{bmatrix}
    \alpha_p&0 \\ 0& \alpha_q \\
\end{bmatrix}R(\theta)')\boldsymbol{u}\\
    = &\ [\boldsymbol{s}_p\ \boldsymbol{s}_q](R(\theta)\begin{bmatrix}
    \beta_p&0 \\ 0& \beta_q \\
\end{bmatrix}R(\theta)')\boldsymbol{u}\\
    \Leftrightarrow\ M^{-1}(\theta)\big|_{pq}= & R(\theta)\begin{bmatrix}
    \beta_p&0 \\ 0& \beta_q \\
\end{bmatrix}R(\theta)' = R(\theta)diag(\beta_p,\ \beta_q)R(\theta)'
\end{aligned}
\end{equation}

for $M^{-1}(\theta)\big|_{pq}$ defining the $M^{-1}$ term restricted within coordinates relative to the fixed orthonormal basis $(\boldsymbol{s}_p\ \boldsymbol{s}_q)$. The expression $M^{-1}(\theta)\big|_{pq}\boldsymbol{x}$ can be expressed as the part of $\boldsymbol{y}$ which is defined over this basis and affected by the change in $\theta$.\\

Expanding $R(\theta)\mathrm{diag} (\beta_p,\beta_q)R(\theta)'$, for $\cos2\theta=\cos^2\theta-\sin^2\theta$ and $\sin2\theta=2\sin\theta\cos\theta$, we have:

\begin{equation}
\begin{aligned}
    &R(\theta)\mathrm{diag}(\beta_p,\beta_q)R(\theta)'\\
    &=\begin{bmatrix}
    \beta_p\cos^2\theta+\beta_q\sin^2\theta & (\beta_p-\beta_q)\cos\theta\sin\theta\\
    (\beta_p-\beta_q)\cos\theta\sin\theta & \beta_p\sin^2\theta+\beta_q\cos^2\theta
    \end{bmatrix}\\
    &=\beta_s I+\beta_d
    \begin{bmatrix}
    \cos 2\theta & \sin 2\theta\\
    \sin 2\theta & -\cos 2\theta
    \end{bmatrix}
\end{aligned}
\end{equation}

such that $\boldsymbol{y}=M(\theta)^{-1}\big|_{pq}\boldsymbol{x}$ is equivalent to:

\begin{equation}
\boldsymbol{y}
=\beta_s \boldsymbol{x}
+ \beta_d
\begin{bmatrix}
\cos(2\theta) & \sin(2\theta)\\
\sin(2\theta) & -\cos(2\theta)
\end{bmatrix} \boldsymbol{x}
\end{equation}

This does two things: it isolates the part of $\boldsymbol{y}$ which is affected by $\theta$, and, given $\boldsymbol{y}$, $\boldsymbol{x}$ are observed and $\beta_s$ is fixed, $\boldsymbol{r} = \boldsymbol{y}-\beta_s \boldsymbol{x}$ can be defined, allowing an avenue to estimating $\theta$. To see how, notice the following must hold:

\begin{equation}
\begin{aligned}
    &\beta_d
\begin{bmatrix}
\cos(2\theta) & \sin(2\theta)\\
\sin(2\theta) & -\cos(2\theta)
\end{bmatrix} \boldsymbol{x}=\beta_d\begin{bmatrix}
x_p & x_q\\
-x_q & x_p
\end{bmatrix}\begin{bmatrix}
\cos(2\theta) \\
\sin(2\theta)
\end{bmatrix}=\boldsymbol{r}\\
\Leftrightarrow
&\begin{bmatrix}\cos(2\theta)\\ \sin(2\theta)\end{bmatrix}
=\frac{1}{\beta_d||\boldsymbol{x}||^{2}}
\begin{bmatrix} x_p & -x_q \\ x_q & \ \ x_p \end{bmatrix}
\boldsymbol{r}
\end{aligned}
\end{equation}

assuming $\beta_d||\boldsymbol{x}||^2\neq0$; and while a unique solution, any LHS satisfying this equation only reflects an optimal angle $\theta$ if $||(\cos(2\theta)\  \sin(2\theta))'||=1$. We therefore have a unique solution in $\theta$ if an only if these conditions apply, though they may be approximated. Then, we can find what that $\theta$ is as follows:\\

\begin{equation}
    \theta=\frac{1}{2}\mathrm{atan2}(\hat {\sin(2\theta)},\hat {\cos(2\theta)})
\end{equation}

unique up to adding $\pi$ - so any optimisation must have in mind a $\theta\in[0,\pi)$. 

\subsubsection{Estimating $\Gamma$}

Estimating $\Gamma$ is far simpler. Under the orthonormality of $S$, it follows from (\ref{shermanmorr}) that:

\begin{equation}
    M^{-1}=I-\sum_{k=1}^K\frac{\gamma_k}{1+\gamma_k}\boldsymbol{s}_k\boldsymbol{s}_k'
\end{equation}

Defining $\beta_k=\frac{\gamma_k}{1+\gamma_k},\ \forall k=1,\ldots,K$, the vector $\boldsymbol{\beta}$ can be estimated in the same way as the $\boldsymbol{\alpha}$ parameters in (\ref{spatecon}).

\end{document}